\documentclass{article}
\usepackage[utf8]{inputenc}
\usepackage{amsmath}
\usepackage{graphicx}
\usepackage{amssymb}
\usepackage{cite}
\usepackage{bm}
\usepackage{amsthm}
\usepackage{subcaption}
\usepackage{hyperref}
\usepackage{authblk}
\usepackage{caption}
\usepackage{subcaption}
\usepackage[toc,page]{appendix}
\usepackage[margin=1in]{geometry}

\usepackage[textwidth=1.7cm,textsize=small]{todonotes}

\title{Continuation of Bianchi Spacetimes Through The Big Bang}

\author{Josh Hoffmann\thanks{Corresponding author: j.hoffmann1@lancaster.ac.uk} } 
\author{David Sloan}
\affil{Department of Physics, Lancaster University, Lancaster, United Kingdom LA1 4YB}

\date{}

\begin{document}

\maketitle

\begin{abstract}
    In this paper we present a framework in which the relational description of General Relativity can be used to smoothly continue cosmological dynamical systems through the Big Bang without invoking quantum gravity effects. Cosmological spacetimes contain as a key dynamical variable a notion of scale through the volume factor $\nu$. However no cosmological observer is ever able to separate their measuring apparatus from the system they are measuring, in that sense every measurement is a relative one and measurable dynamical variables are in fact dimensionless ratios. This is manifest in the identification of a scaling symmetry or ``Dynamical Similarity" in the Einstein-Hilbert action associated with the volume factor. By quotienting out this scaling symmetry, we form a relational system defined on a contact manifold whose dynamical variables are decoupled from scale. When the phase space is reduced to shape space, we show that there exist unique solutions to the equations of motion that pass smoothly through the initial cosmological singularity in flat FLRW, Bianchi I and Quiescent Bianchi IX cosmologies.
\end{abstract}

\tableofcontents

\clearpage

\section{Introduction}

One of the biggest conceptual and practical problems in contemporary cosmology is that of singularities in General Relativity (GR). The Hawking-Penrose theorems \cite{1970RSPSA.314..529H} establish that a wide class of spacetimes are geodesically incomplete.
%In other words, singularities occur generically.%
%
Without going into the details of measures on a space of theories, this is generally taken to mean that maximally extended space-times without any singularities are a set of measure zero. 
Solutions to the Einstein Field Equations (EFE) provide the metric structure and manifold geometry at each point in spacetime, when the EFE become indeterminate, the spacetime structure cannot be defined. In this sense there is a boundary beyond which physics cannot be deterministically continued which we know as a singularity. Typically this is characterised by some component of the Ricci tensor growing arbitrarily large \cite{CLARKE1985127}, although this is not always the case. The most comprehensive understanding of singularities is provided by the Hawking-Penrose theorems, in which geodesic incompleteness necessitates a spacetime singularity, there will be a boundary beyond which a null or timelike geodesic terminates after finite 
%proper time%
% null geodesics have no proper time
affine length
\cite{1970RSPSA.314..529H,Hawking:1973uf}. This poses a problem for General Relativity. Many physical solutions are bound to contain a point region beyond which physics cannot be continued in a deterministic manner. In particular, cosmological solutions to GR contain a Big Bang singularity, characterised by the vanishing of the scale factor and the infinite growth of the Hubble parameter. Study of the Big Bang has been at the centre of contemporary cosmology research \cite{2015imct.book.....L}. In most modern literature, attempts to resolve spacetime singularities invoke a quantisation of gravity, be it perturbatively through string theory \cite{2008APS..APR.B7003H,natsuume2001singularity, Gibbons_1995} or non-perturbatively through Loop Quantum Gravity \cite{Li:2023dwy,Gambini:2013hna,Gambini:2022hxr,Ashtekar:2006rx,Bojowald:2001xe,Bojowald:2005epg}. 
These resolutions can be broadly fit into two camps. In the first camp the singularity is avoided dynamically through, e.g. the introduction of a new type of matter, or repulsive force at small scales. In the second, the space-time geometry is replaced by a quantum counterpart such as a quantum foam. In both cases the resolution requires that GR be `corrected' in the neighbourhood of the singularity. The need for these corrections informs two key aspects in the study of quantum gravity. In the first the necessity of replacing GR by a quantum counterpart is underlined by the presence of singular points at which the theory needs to be corrected. In turn this means that that such points are the focus of where quantum theories of gravity should provide observable deviations from GR. 
In this work, we present an approach to the resolution of spacetime singularities in an entirely classical manner, appealing to the relationalist framework of Shape Dynamics (SD) \cite{Mercati:2014ama,Barbour:2011dn}. A key dynamical variable in the cosmological sector of GR is the scale factor $a$. The scale factor however, is not a physical observable. One never measures (or more precisely, is not able to measure) the size of the universe $a$, only the size compared to some reference scale $a/a_{\text{ref}}$. Any measurement of the size of universe is a relative one, as our measuring apparatus is indeed part of the system we are attempting to measure. The relationalist framework laid out by Barbour and Bertotti \cite{Barbour:1982gha} seeks to track only the intrinsic change of a dynamical system. This has formed the foundation of the Shape Space, and recently, Dynamical Similarity \cite{Sloan_2018} approach to cosmology. In this paper, we synthesize recent work on the Shape Dynamics formalism \cite{Koslowski:2016hds,Sloan:2019wrz} and Dynamical Similarity to show how, starting with the ADM description of Hamiltonian GR, one can move formally to a description of cosmology with equivalent physical dynamics that never makes reference to a notion of absolute scale. In this description the dynamical system of physical observables defined on the phase space manifold remains well defined through the initial cosmological singularity and in this sense it is resolved classically. In this description the principle mathematical objects, namely the Hamiltonian and contact structure, remain finite and well defined, which hints at a rigorous route to quantistaion. 

This classical resolution is possible because we relax the requirement that our physical system consists of a pseudo-Riemannian geometry at all times. With the elimination of scale from our system we form a physical entity which, away from the singularity, can always be embedded within a pseudo-Riemannian framework through the additional assignment of scale. However, while this assignment breaks down at the singularity, crucially the integrability of the equations of motion does not. Therefore within this framework the singularity is not an endpoint of physics, but rather a point at which the pseudo-Riemannian description in terms of a three dimensional spatial manifold with volume is no longer appropriate. 
The paper is structured as followed. In section \ref{contactsect} we detail the mechanics of contact manifolds, which will be used to describe the phase space of dynamical systems later on. In section \ref{Dynsect} we describe dynamical similarities, which deal with the  fact that, physically measurable quantities in cosmology are insensitive to the overall volume $\nu \sim a^3$ represents a redundancy in the description of the system. The formalism of dynamical similarities will be used to quotient out the redundancy of the scale factor, moving from the even-dimensional symplectic manifold description in canonical ADM General Relativity, to a system defined on an odd-dimensional contact manifold with no reference to the scale factor. This constitutes the ``Contact Reduction'' formalism. Section \ref{ADMsect} gives a brief overview of the ADM formalism and the homogeneous spacetimes with which we will be concerned. In sections \ref{FLRWsect} and \ref{Bianchisect} we show that flat FLRW and Bianchi Cosmologies can be contact-reduced to form dynamical systems that are fully decoupled from scale. In section \ref{shapesect} we show that the equations of motion of the new contact-reduced systems, when projected to Shape Space satisfy the Picard-Lindelöf theorem in a neighbourhood of the initial singularity. Thus there exists unique solutions to the equations of motion that pass smoothly through the singularity. In section \ref{numsect} we provide numerical solutions to the equations of motion for all of the cosmologies considered in this paper. In section \ref{dissect} we discuss the results and how this formalism fits into the wider literature of attempts to resolve spacetime singularities, particularly with respect to quantisations of GR.

\subsection{Contact Mechanics}
\label{contactsect}

Typically Lagrangian systems in physics admit an equivalent Hamiltonian description, defined on a symplectic manifold $(M,\omega)$. In GR it is possible, although not trivial, to find such a Hamiltonian description. It is non-trivial because GR does not admit a natural time parameterisation which is needed for the Hamiltonian description. Choosing a particular time parameterisation breaks general covariance of the theory. However, by assuming the spacetime manifold to be globally hyperbolic, so that is has the topology $\mathcal{M} = \mathbb{R}\times \Sigma$, where $\Sigma$ is a 3-dimensional orientable manifold. The globally hyperbolic spacetime manifold can be foliated by spacelike hypersurfaces $\Sigma_t$ labelled by a parameter $t\in\mathbb{R}$. This is the basis for the ADM formalism \cite{PhysRev.116.1322,Jha:2022svf}, which provides a Hamiltonian description of GR by considering the evolution of these hypersurfaces with the parameter ${t}$.

The tools of symplectic geometry can be used to write Hamilton's equations of motion in a coordinate invariant way, which is obviously useful in the context of the ADM formalism as we will be interested in dynamics on spatial 3-manifolds $\Sigma_t$.
A symplectic manifold $(M, \omega)$ is a $2n$-dimensional manifold $M$ along with a non-degenerate, closed 2-form $\omega$ called the symplectic structure. If $H: M \rightarrow \mathbb{R}$ is a time-independent Hamiltonian function on $M$, then Hamiltons equations of motion are generated by the symplectic form
\begin{equation}
\label{X_H}
    i_{X_H} \omega = dH
\end{equation}
Henceforth we shall assume all Hamiltonian functions in this paper to be time independent, but the standard results can be trivially extended to time-dependent Hamiltonians \cite{10.3390/e20040247, deLeon:2020hnm}.
Typically, we work within a context where the even-dimensional manifold in question is a cotangent bundle $T^*Q$ of the configuration space $Q$. We may always find local Darboux coordinates $(q_i,p^i)$ on the cotangent bundle such that there is a natural symplectic structure \cite{1978mmcm.book.....A,2017AnPhy.376...17B}
\begin{equation}
    \omega = dq_i \wedge dp^i
\end{equation}
and the unique Hamiltonian vector field satisfying \ref{X_H} is 
\begin{equation}
    X_H = \frac{\partial H}{\partial p^i}\frac{\partial}{\partial q_i} - \frac{\partial H}{\partial q_i}\frac{\partial}{\partial p^i}
\end{equation}
The solutions to Hamiltons equations of motion are then integral curves of the Hamiltonian vector field
\begin{equation}
    \dot{q}_i = \frac{\partial H}{\partial p^i}, \quad \dot{p}^i = -\frac{\partial H}{\partial q_i}
\end{equation}
The natural symplectic form admits a Poisson bracket structure on the cotangent bundle manifold, such that the time evolution of any smooth function $F(q_i,p^i) \in C^\infty(M)$ is given by
\begin{equation}
    \frac{dF}{dt} = X_H[F] = \{F,H\} = \omega(X_F,X_H)
\end{equation}

Symplectic geometry is concerned with even dimensional manifolds, but in this paper we will make use of symmetry-reduced models where one degree of freedom, namely the scale factor $\nu$ has been removed. The manifolds on which such systems live are odd-dimensional, so we will need the mechanics of contact structures \cite{10.3390/e20040247,deLeon:2020hnm,bravetti2020invariant}.

Like a symplectic manifold, a contact manifold $(M,\eta)$ is a pair consisting of an odd-dimensional manifold $M$ with $\text{dim}(M) = 2n+1$ and a non-degenerate 1-form $\eta$ called the contact form, satisfying
\begin{equation}
    \eta \wedge (d\eta)^n \neq 0 
\end{equation}
Physical paths of the system on the contact manifold are again integral curves of a unique Hamiltonian vector field $X_H$ satisfying
\begin{equation}
\label{Contact_XH}
    -i_{X_H}\eta = dH - \mathcal{R}(H)\eta, \quad i_{X_H}\eta = H
\end{equation}

where $\mathcal{R}$ is the Reeb vector field. This is a unique vector field satisfying 
\begin{equation}
    \eta(\mathcal{R}) = 1, \quad d\eta(\mathcal{R}) = 0
\end{equation}
Just as in the symplectic manifold case, if $M = T^*Q \times \mathbb{R}$, where $T^*Q$ is the cotangent bundle, it is always possible to find local Darboux coordinates in which the contact form is
\begin{equation}
    \eta = -dS + p^i dq_i
\end{equation}
where $(q_i,p^i,S)$ are local coordinates on $M$. In these local Darboux coordinates, the Reeb vector field is
\begin{equation}
    \mathcal{R} = -\frac{\partial}{\partial S}
\end{equation}
and the unique Hamiltonian vector field satisfying \ref{Contact_XH} is 
\begin{equation}
    X_H = \frac{\partial H}{\partial p^i}\frac{\partial}{\partial q_i} -\left(\frac{\partial H}{\partial q_i} + p^i\frac{\partial H}{\partial S}\right)\frac{\partial}{\partial p^i} + \left(p^i\frac{\partial H}{\partial p^i} - H\right)\frac{\partial}{\partial S}
\end{equation}
The integral curves satisfied by physical paths on the manifold are then
\begin{equation}
    \label{Conact_eom}
    \begin{aligned}
        &\dot{q}_i = \frac{\partial H}{\partial p^i} \\
        &\dot{p}^i = -\frac{\partial H}{\partial q_i} - p^i\frac{\partial H}{\partial S} \\
        &\dot{S} = p^i\frac{\partial H}{\partial p^i} - H
    \end{aligned}
\end{equation}
The contact Hamiltonian equations of motion are a natural generalisation of the standard symplectic Hamilton equations of motion. If the contact Hamiltonian is independent of $S$, one recovers the symplectic equations of motion for $(q_i(t),p^i(t))$.
The time evolution of any smooth functions $F \in C^\infty(M)$ is generated by a natural Poisson bracket-structure admitted by the Hamiltonian vector field
\begin{equation}
\label{dt}
    \frac{d F}{dt} = X_H[F] = -H\frac{\partial F}{\partial S} + \{F,H\}_{(q_i,p^i)} + p^i\{F,H\}_{(S,p^i)}
\end{equation}

It is well known that in the symplectic case, the Hamiltonian is conserved along its flow, expressing energy conservation of the system. Consequently Liouville's theorem holds and the Lie drag of the symplectic volume form is zero along the physical paths of the system. 
\begin{equation}
    \mathcal{L}_{X_H}\omega = 0
\end{equation}
For a contact system this is not necessarily the case. From equation \ref{dt} one can see that the total time derivative of the Hamiltonian is 
\begin{equation}
    \frac{dH}{dt} = -H\frac{\partial H}{\partial S}
\end{equation}
which is in general non-zero unless the Hamiltonian is not a function of the global phase space coordinate $S$ or $H = 0$ everywhere.

Just as how in the symplectic case, one can just as equally start with a Lagrangian function $L: TQ \rightarrow \mathbb{R}$
such that the physical paths on the manifold satisfy the Euler-Lagrange equations
\begin{equation}
    \frac{d}{dt}\left(\frac{\partial L}{\partial \dot{q}_i}\right) - \frac{\partial L}{\partial q_i} = 0
\end{equation}
if and only if they are stationary paths of the action
\begin{equation}
    S = \int L(q_i(t),\dot{q}_i(t)) dt
\end{equation}
and use a Legendre transformation $FL:TQ\rightarrow T^*Q$ to go between the Lagrangian and symplectic Hamiltonian \cite{silva2001lectures} $H(q,p) = p^i\dot{q}_i - L(q_i, \dot{q}_i)$, assuming that the Lagrangian is regular and convex.

In the case of contact mechanics, one can start with a Lagrangian function on an odd-dimensional $(2n+1)$ manifold $L : TQ\times \mathbb{R}\rightarrow \mathbb{R}$, and define a contact form $\eta_L$ using the canonical endomorphism  $E$ on $TQ$
\begin{equation}
    \begin{aligned}
    \label{eta_L}
        \eta_L &= -dS + E^*(dL) \\
        &= -dS + \frac{\partial L}{\partial \dot{q}_i}dq_i
    \end{aligned}
\end{equation}
where $(q_i,\dot{q}_i)$ are the bundle coordinates on $TQ$ and $S$ is the coordinate on $\mathbb{R}$
On any contact manifold, there exits a unique Reeb vector field $R$ satisfying
\begin{equation}
    i_R\eta_L = 1, \quad i_R d\eta_L = 0
\end{equation}
Given the contact form \ref{eta_L}, the Reeb vector field expressed in the local coordinates is
\begin{equation}
    R_L = -\frac{\partial}{\partial S} + W^{ij}\frac{\partial^2}{\partial \dot{q}_j\partial S} \frac{\partial}{\partial \dot{q}_i}
\end{equation}
where $W^{ij}$ is the inverse Hessian
\begin{equation}
    W_{ij} = \frac{\partial^2L}{\partial\dot{q}_j\partial\dot{q}_j}
\end{equation}
which is assumed to be regular. To achieve the Herglotz equations of motion, which are generalised Euler-Lagrange equations, one takes the physical paths on the manifold to be the integral curves of a unique Lagrangian vector field $\xi_L$ satisfying
\begin{equation}
    I_L(\xi_L) = dE_L - \left[R_LE_L + E_L\right]\eta_L
\end{equation}
where $I_L$ is a the vector bundle isomporphism $I_L : T(TQ\times\mathbb{R})\rightarrow T^*(TQ\times\mathbb{R})$ admitted by the contact form $\eta_L$
\begin{equation}
    I_L(v) = i_v(d\eta_L) + (i_v\eta_L)\eta_L
\end{equation}
and $E_L$ is the scalar energy of the system defined by 
\begin{equation}
    E_L = \dot{q}_i\frac{\partial L}{\partial\dot{q}_i} - L
\end{equation}
the integral curves of $\xi_L$ satisfy the Herglotz equations of motion
\begin{equation}
\label{Herglotz_eom}
    \frac{d}{dt}\left(\frac{\partial L}{\partial \dot{q}_i}\right) - \frac{\partial L}{\partial q_i} = \frac{\partial L}{\partial S}\frac{\partial L}{\partial \dot{q}_i} 
\end{equation}
which reduce the standard Euler-Lagrange equations for $q_i\dot{q}_i$ in the case $L(q_i,\dot{q}_i,S) = L(q_i,\dot{q}_i)$
The integral curves satisfy the Herglotz equations of motion \ref{Herglotz_eom} if and only if they are stationary paths of the action.

\begin{equation}
    S = \int L(q_i,\dot{q}_i,S) dt
\end{equation}

As one would expect, this two geometric descriptions of the Euler-Lagrange and Hamiltonian equations of motion are equivalent. Given a contact form $\eta = -dz + p^idq_i$ on $T^*Q\times \mathbb{R}$ and a Legendre transform $FL:TQ\times \mathbb{R} \rightarrow T^*Q\times\mathbb{R}$ of a Herglotz Lagrangian function $L(q_i,\dot{q}_i,z)$ defined as
\begin{equation}
    FL(q_i,\dot{q}_i,z) = (q_i,p^i,z), \quad p^i = \frac{\partial L}{\partial\dot{q}_i}
\end{equation}
the pullback of $\eta$ by $FL$ is explicitly
\begin{equation}
    FL^*(\eta) = -dz + \frac{\partial L}{\partial \dot{q}_i}dq_i = \eta_L
\end{equation}
thus a Legendre transformation moves between the contact Hamiltonian and contact Lagrangian equations of motion.

\section{Contact Reduction Theory}
\subsection{Dynamical Similarity}
\label{Dynsect}

Here we present how scale invariance can be used to reduce a theory from a symplectic manifold to a contact manifold. For a formal mathematical exposition see \cite{bravetti2023scaling}. Scale invariance is an important aspect of the analysis of dynamical systems in cosmology. In any cosmological measurement, the scale factor is never directly observable as we are only able to infer relative changes. Consider for example the Kepler problem as in \cite{sloan2021scale}. We consider the Lagrangian of two unit mass particles interacting through Newtonian gravity

\begin{equation}
    L = \frac{1}{2}\dot{r}^2 + \frac{1}{2}r^2\dot{\theta}^2 + \frac{G}{r}
\end{equation}
where $r$ is the separation in the centre of mass frame. There exists a coordinate transformation that moves between indistinguishable solutions of the equations of motion. In particular this coordinate transformation is a multiplicative rescaling by a non-zero constant $\lambda$
\begin{equation}
    r \rightarrow \lambda r, \quad t \rightarrow \lambda^{\frac{3}{2}}t
\end{equation}
Under this rescaling, the action
\begin{equation}
    S = \int L(r,\dot{r},\dot{\theta}) dt
\end{equation}
transforms also by a multiplicative rescaling $S \rightarrow \lambda^\frac{1}{2}S$ and hence the equations of motion whose solutions are critical paths of the action $\delta S = 0$, remain unchanged. This is an example of a dynamical similarity \cite{Sloan_2018}, which is a scaling symmetry of the dynamical system.  As we will see in sections 2.2 and 2.3, Lagrangians of relevant cosmological models exhibit a dynamical similarity associated with the scale factor. In this section (2.1) we show how the identification of a dynamical similarity can be used to quotient out a redundant symmetry of the symplectic phase space, and define a system with equivalent dynamics on a contact manifold phase space.

Consider a symplectic system with Lagrangian $L(q_i,\dot{q}_i): TQ\rightarrow \mathbb{R}$ and action
\begin{equation}
\label{Sym_Action}
    S = \int L(q_i,\dot{q}_i) dt
\end{equation}
integrated over physical paths $\gamma$ of the tangent bundle manifold $TQ$ of a configuration space $Q$, parameterised by time over some interval $I\subseteq \mathbb{R}$. The physical cuvres $\gamma$ on $TQ$ are tangent lifts of curves $\bar{\gamma}$ on the configuration space. In this paper we will consider Lagrangians that have a configuration space scaling symmetry (CSSS). We define a CSSS to be a vector field $\mathbf{\bar{D}}$ on $Q\times I$ such that
\begin{itemize}
    \item For all curves $\bar{\gamma} : I \rightarrow Q$ whose tangent lifts are stationary paths of the action \ref{Sym_Action}, $\mathbf{\bar{D}}\bar{\gamma}$ are also stationary paths of the action.
    \item $\mathbf{D}L = \Lambda L$, where $\Lambda \in \mathbb{R}$ and $\mathbf{D}$ is the tangent lift of the CSSS
\end{itemize}
The scaling symmetry $\bar{\mathbf{D}}$ moves between curves on configuration space $Q$, which represent indistinguishable solutions of the equations of motion as the Lagrangian scales by a multiplicative constant $\Lambda$ which we call the degree of the CSSS. The CSSS is defined only up to a non-zero scalar factor, which we may fix by demanding that $\mathbf{D}$ preserves the Lie drag of the Lagrange one-form 
\begin{equation}
    \mathcal{L}_{\mathbf{D}}\mu_L = \mu_L, \quad \mu_L = \frac{\partial L}{\partial \dot{q}_i}dq_i
\end{equation}
Consider a scaling symmetry of one of the configuration space variables $x\in Q$ and possibly of the time coordinate $t \in I$ and write the new coordinates under the scaling symmetry as 
\begin{equation}
    (\bar{x},\bar{t}) = (Ax,Bt) = (\bar{\mathbf{D}}x,\bar{\mathbf{D}}y), \quad \bar{\mathbf{D}} = D_x\partial_x + D_t\partial_t
\end{equation}
Thus the CSSS can be written locally w.l.o.g. as 
\begin{equation}
    \bar{\mathbf{D}} = Ax\frac{\partial}{\partial x} + Bt\frac{\partial}{\partial t}
\end{equation}
which has tangent lift acting on the Lagrangian $\mathbf{D}L = \Lambda L$
\begin{equation}
    \mathbf{D} = Ax\frac{\partial}{\partial x} + (A-B)\dot{x}\frac{\partial}{\partial\dot{x}} - B\dot{q}_i\frac{\partial}{\partial\dot{q}_i} + Bt\frac{\partial}{\partial t}
\end{equation}
where $q_i\in Q$ are the configuration space coordinates unaffected by the scaling symmetry. 
We next make a coordinate and lapse transformation 
\begin{equation}
    \rho = x^{\frac{1}{A}}, \quad d\tau = \rho^{-B}dt
\end{equation}
under which the CSSS is isochronal $(B = 0)$ and has tangent lift
\begin{equation}
    \mathbf{D} = A\left(x\frac{\partial}{\partial x} + x'\frac{\partial}{\partial x'}\right) \sim x\frac{\partial}{\partial x} + x'\frac{\partial}{\partial x'}, \quad \text{Normalising } A = 1
\end{equation}
where $'$ denotes derivatives w.r.t. the new time coordinate $\tau$. We know show that given a Lagrangian function $L:TQ\rightarrow \mathbb{R}$ with a CSSS $\mathbf{D}$ of degree 1, it is possible to construct a Herglotz Lagrangian $L^H:TQ\times\mathbb{R}\rightarrow\mathbb{R}$ that describes the same dynamics on $TQ/\mathbf{D}$.
Firstly, consider the converse. Let $L^H(q,\dot{q},S)$ be a Herglotz Lagrangian, and define 
\begin{equation}
    L = e^\rho\left(L^H + \dot{\rho}S\right), \quad \rho = -\frac{\partial L^H}{\partial S}
\end{equation}
Since $L^H$ is a Herglotz Lagrangian, the physical paths on the manifold satisfy eq \ref{Herglotz_eom}.
Thus one can show that the original tangent bundle coordinates $(q,\dot{q})$ satisfy the Euler-Lagrange equations generated by $L$.
\begin{equation}
    \begin{aligned}
        \frac{d}{dt}\left(\frac{\partial L}{\partial q}\right) - \frac{\partial L}{\partial q} &= \frac{d}{dt}\left(\frac{\partial L^H}{\partial\dot{q}}\right) - \frac{\partial L^H}{\partial q} + \dot{\rho}\frac{\partial L^H}{\partial\dot{q}} \\
        &= \frac{d}{dt}\left(\frac{\partial L}{\partial \dot{q}} - \frac{\partial L^H}{\partial q}\right) - \frac{\partial L^H}{\partial S}\frac{\partial L^H}{\partial \dot{q}} = 0
    \end{aligned}
\end{equation}
Thus the dynamics of all $(q,\dot{q})\in TQ$ generated by $L$ and $L^H$ are equivalent. Lastly make a change of coordinate $x = e^\rho$, under which the Lagrangian becomes
\begin{equation}
    \label{xL}
    L = xL^H + \dot{x}S
\end{equation}
The Lagrangian \ref{xL} clearly has a scaling symmetry of degree 1 given by
\begin{equation}
    \label{D1}
    \mathbf{D} = x\partial_x + \dot{x}\partial_{\dot{x}}
\end{equation}
Now consider the converse, let $L:TQ\rightarrow \mathbb{R}$ be a Lagrangian with the scaling symmetry \ref{D1} of degree 1.
By the definition of $\mathbf{D}$ we have $\mathbf{D}L = L$ thus
\begin{equation}
    x\frac{\partial L}{\partial x} + \dot{x}\frac{\partial L}{\partial\dot{x}} = L
\end{equation}
We identify the partial derivatives as 
\begin{equation}
    \label{Herglotz_L}
    L^H = \frac{\partial L}{\partial x}, \quad S = \frac{1}{x}i_{\mathbf{D}}\mu_L = \frac{\partial L}{\partial\dot{x}}
\end{equation}
where $L^H(q,\dot{q},S):TQ\times\mathbb{R}$ can be shown to be a Herglotz Lagrangian with equivalent dynamics to $L$.
The physical paths satisfy the Euler-Lagrange equations generated by $L$, which can be written in terms of $L^H$.
\begin{equation}
    \begin{aligned}
        &\frac{d}{dt}\left(\frac{\partial L}{\partial \dot{q}}\right) - \frac{\partial L}{\partial q} = 0 \\
        & \implies x\frac{d}{dt}\left(\frac{\partial L^H}{\partial\dot{q}}\right) - x\frac{\partial L^H}{\partial x} -L^H\delta_{x,q} + \dot{S}\delta_{\dot{x},\dot{q}} -x\frac{\partial L^H}{\partial S}\frac{\partial L^H}{\partial \dot{q}} = 0 \\
        &\implies \frac{d}{dt}\left(\frac{\partial L^H}{\partial \dot{q}}\right) - \frac{\partial L^H}{\partial q} - \frac{\partial L^H}{\partial S}\frac{\partial L^H}{\partial \dot{q}} = 0 
    \end{aligned}
\end{equation}
Thus the physical paths also satisfy the Herglotz equations of motion generated by $L^H$ on the symmetry reduced manifold $TQ/\mathbf{D}$.

\subsection{Cosmological Space-times}
\label{ADMsect}

In this section we present a brief review of the ADM formalism of General Relativity and cosmological spacetimes of interest in this paper, namely Bianchi (I and IX) and flat FLRW. We refer the reader to \cite{Jha:2022svf} for more detail. 

Within the framework of General Relativity, spacetime is assumed to be a 4-dimensional Lorentzian differentiable manifold $\mathcal{M}$ with metric $g_{\mu\nu}$. The manifold is assumed to be globally hyperbolic, so that it has the topological structure of $\mathbb{R}\times\Sigma$ where $\Sigma$ is a spatial 3-manifold. Since $\mathcal{M}$ is globally hyperbolic, it can be foliated by such spacelike hypersurfaces.
The full spacetime metric may be decomposed as 
\begin{equation}
    g_{\mu\nu} = \begin{pmatrix}
        N_iN^i - N^2 & N_j \\
        N_i & \gamma_{ij}
    \end{pmatrix}
\end{equation}
where the shift vector $N^i$, lapse function $N$ and spatial 3-metric $\gamma_{ij}$ fully determine the 4-dimensional spacetime metric $g_{\mu\nu}$. With the 3+1 foliation of spacetime described above, one may start with the Einstein-Hilbert action and decompose it into dynamical variables defined on the 3-geometry

\begin{equation} 
\label{S_EH}
\begin{aligned}
    S &= \int d^4x \sqrt{-g}R \\
    &= \int d^4x N\sqrt{\gamma}\left(^3R(\gamma) + K^2 + K_{ij}K^{ij} - 2\nabla_n K -\frac{2}{N}D^iD_i N\right)
\end{aligned}
\end{equation}
where $^3R$ is the Ricci scalar associated with the spatial 3-metric $\gamma_{ij}$, $K_{ij}$ is the extrinsic curvature of the spatial spatial slice $\Sigma_t$ given by
\begin{equation}
    K_{ij} = -\gamma^\sigma_i\gamma^\rho_j\nabla_\sigma n_\rho = \frac{1}{2N}\left(D_iN_j + D_jN_i - \dot{\gamma}_{ij}\right)
\end{equation}
$K = \gamma_{ij}K^{ij}$ is the extrinsic curvature scalar, $D_i$ are the spatial covariant derivatives and $\nabla_n K$ is the covariant derivative along the normal vector to the spatial 3-manifold $n^\mu$, normalised to $n^\mu n_\mu = -1$. It is important to note that the objects $\gamma_{ij}$, $K_{ij}$ and $D_i$ are non-zero only on the spatial 3-manifold and therefore only have Latin subscripts. The action in eq. \ref{S_EH} is equivalent to 

\begin{equation}
\label{S}
    S = \int d^4x N\sqrt{\gamma}\left[^3R(\gamma) + K_{ij}K^{ij}-K^2\right]
\end{equation}
up to boundary terms that have no effect on the equations of motion as their variation can be set to zero. Thus we have the Einstein-Hilbert Lagrangian 
\begin{equation}
    \mathcal{L} = N\sqrt{\gamma}\left[^3R(\gamma) + K_{ij}K^{ij}-K^2\right]
\end{equation}
from which we may find the conjugate momenta of the dynamical variables $\{N,N^i,\gamma_{ij} \}$. Firstly one sees immediately that the lapse $N$ and shift vector $N^i$ have no conjugate momenta as their velocities do not enter into the Lagrangian. We will see that the lapse and shift vector actually turn out to be non-dynamical variables, as variation of the action with respect to them results in constraint relations which are fundamental to the Hamiltonian description of GR. On the other hand, the conjugate momenta to the spatial 3-metric is given by

\begin{equation}
    \pi^{ij} = \frac{\partial \mathcal{L}}{\partial \dot{\gamma}_{ij}} = \sqrt{\gamma}\left(K\gamma^{ij}- 
    K^{ij}\right)
\end{equation}

The Hamiltonian density can then be obtained from a Legendre transform $\mathcal{H} = \pi^{ij}\dot{\gamma}_{ij} - \mathcal{L}$, giving the Hamiltonian
\begin{equation}
\label{H_pre}
    H = \int_{\Sigma_t}d^3x \, \mathcal{H} = \int_{\Sigma_t}d^3x \, N\left[\frac{1}{\sqrt{\gamma}}\left(\pi^{ij}\pi_{ij} - \frac{1}{2}\pi^2\right)-\sqrt{\gamma}^3R(\gamma)\right] + 2\int_{\Sigma_t}d^3x \, \pi^{ij}D_iN_j
\end{equation}
The term in the second integral of eq \ref{H_pre} can be written in terms of a spatial covariant derivative of $\pi^{ij}$ up to a boundary term which we may neglect. Thus the Hamiltonian that is conventionally presented in the ADM literature is 
\begin{equation}
    H = \int_{\Sigma_t}d^3x \, N\left[\frac{1}{\sqrt{\gamma}}\left(\pi^{ij}\pi_{ij} - \frac{1}{2}\pi^2\right)-\sqrt{\gamma}^3R(\gamma)\right] - 2\int_{\Sigma_t}d^3x \, N^iD^j\pi_{ij}
\end{equation}

As usual, the physical paths on the solution space are the stationary paths of the action. Requiring the variation of the action \ref{S} with respect to the lapse and shift vector produces
\begin{equation}
\label{H_cons}
    \sqrt{\gamma}^3R(\gamma) = K_{ij}K^{ij}-K^2 = \frac{1}{\sqrt{\gamma}}\left(\pi^{ij}\pi_{ij}-\frac{1}{2}\pi^2\right) = 0
\end{equation}
\begin{equation}
\label{D_cons}
    D^i\pi_{ij} = 0
\end{equation}
respectively. These are constraint equations which must be satisfied by solutions to the equations of motion. The first constraint eq \ref{H_cons} is known as the Hamiltonian constraint and eq \ref{D_cons} is the diffeomorphism constraint. The expression of the ADM Hamiltonian \ref{H_pre} can be written in terms of the constraints by defining 
\begin{equation}
\label{HN}
    \mathbb{H}[N] = \int_{\Sigma_t}d^3x \, N\left[\frac{1}{\sqrt{\gamma}}\left(\pi^{ij}\pi_{ij} - \frac{1}{2}\pi^2\right)-\sqrt{\gamma}^3R(\gamma)\right] \approx 0
\end{equation}
\begin{equation}
\label{DN}
    \mathbb{D}[N^i] = - 2\int_{\Sigma_t}d^3x \, N^iD^j\pi_{ij} \approx 0
\end{equation}
Where $\approx 0$ denotes zero for solutions of the equations of motion which lie on the spacelike hypersurfaces $\Sigma_t$. Thus the Hamiltonian is written as
\begin{equation}
    H = \mathbb{H}[N] + \mathbb{D}[N^i] \approx 0
\end{equation}
In this sense the lapse function and shift vector are non-dynamical Lagrange multipliers. The spatial 3-metric $\gamma_{ij}$ and conjugate momentum $\pi^{ij}$ satisfy Hamiltons equations of motion

\begin{equation}
    \dot{\gamma}_{ij} = \frac{\partial \mathcal{H}}{\partial \pi^{ij}}, \quad \pi^{ij} = -\frac{\partial \mathcal{H}}{\partial\gamma^{ij}}
\end{equation}

\subsection{Homogeneous Cosmologies}
In this paper we will be concerned with spatially homogeneous cosmological spacetimes. The isomotries of a spatially homogeneous but not necessarily isotropic spacetime are captured by the Bianchi classification \cite{Jha:2022svf,Landau:1975pou,MONTANI_2008}, which we will describe briefly here. A spatially homogeneous spacetime $M = \mathbf{R}\times\Sigma$ is one for which there exists a Lie group $G$ such that for any point $p$ on the spatial 3-manifold, every point $q\in \Sigma_t$ lies in the group orbit of $p$. In other words, $\forall p,q \in \Sigma_t$ there exists a unique $g\in G$ such that $q = g(p)$. the group isomotries on the spatial manifold have 3-independent Killing vector fields $\xi_a$ as their infinitesimal generators satisfying the commutation relation

\begin{equation}
    [\xi_a,\xi_b] = C^c_{ab}\xi_c
\end{equation}
where $C^c_{ab}$ are the structure constants of the Lie group $G$. Up until this point we have worked in a coordinate basis $e_i = \partial_i$, $e^i = dx^i$. Note that where the spatial slice is non-compact, we replace $\Sigma_t$ with a fiducial cell - a compact region within the spatial slice which encodes all information due to homogeneity. The existence of the Killing vector fields on a homogeneous spacetime implies the existence of a non-coordinate basis ${e_a}$ satisfying the same commutation relation as the Killing vector field
\begin{equation}
    [e_a,e_b] = C^c_{ab}e_c
\end{equation}
that is invariant under the group action generated by the Killing vector fields. The Bianchi classification is thus the identification of all (inequivalent) Lie groups of dimension 3. This classification is of course well known and has been studied extensively in the literature. Type A Bianchi cosmologies are those which admit a Hamiltonian description and are thus the only ones we will be concerned with in this paper. Their structure constants may be decomposed as 
\begin{equation}
    C^c_{ab} = \epsilon_{abd}C^{dc}
\end{equation}
where $\epsilon_{abd}$ is the Levi-Civita symbol and $C^{dc}$ is a symmetric matrix, fully characterised by its eigenvalues $n_1,n_2,n_3 \in \{-1,0,1\}$. In this paper we will be concerned with the Bianchi I and Bianchi IX models which are described by the eigenvalues $n_i = 0$ and $n_i = 1$ respectively.

In the case of a Bianchi I cosmology, the structure constants are all identically zero and the invariant basis coincides with the coordinate basis. The Hamiltonian comstraint in this case can be written as
\begin{equation}
    \mathbb{H}[N] = \int_{\Sigma_t} d^3x N\left[\frac{1}{\sqrt{\gamma}}\left(\pi^{ij}\pi_{ij} - \frac{1}{2}\pi^2\right)\right]
\end{equation}
as the 3-Ricci scalar is everywhere zero. The invariant basis coinciding with the coordinate basis also means that the  diffeomorphism constraint is satisfied trivially.
\begin{equation}
  \mathbb{D}[N^a] = -2 \int_{\Sigma_t} d^3x N^a D^b\pi_{ab}(t) = 0
\end{equation}
Thus the Bianchi I Hamiltonian can be written as 
\begin{equation}
    H_{BI} = \frac{n}{\sqrt{\gamma}}\left(\frac{1}{2}\pi^{ab}\pi_{ab}-\frac{1}{2}\pi^2\right)
\end{equation}
where $n = \int_{\Sigma_t}d^3xN$ is the lapse function spatially integrated over the fiducial cell. The integration over the fiducial cell takes us from field theory of GR with an infinite number of degrees of freedom (although a finite number per point), to a particle theory with a finite number of degrees of freedom. This is a non-dynamical Lagrange multiplier and so can be set to unity without loss of generality. The remaining degrees of freedom can be used to work in the Taub gauge where the metric and it's conjugate momentum are diagonal. In particular we choose
\begin{equation}
\label{gamma}
    \gamma_{ab} = \alpha \,\text{diag}\left(e^{-x+\frac{y}{\sqrt{3}}}, e^{x+\frac{y}{\sqrt{3}}}, e^{-\frac{2}{\sqrt{3}}y}\right)
\end{equation}
\begin{equation}
\label{pi}
    \pi^{ab} = \alpha^{-1} \, \text{diag}\left[\left(-\frac{p_x}{2}+\frac{p_y}{2\sqrt{3}}+D\right)e^{x-\frac{y}{\sqrt{3}}},\left(\frac{p_x}{2}+\frac{p_y}{2\sqrt{3}}+D\right)e^{-x-\frac{y}{\sqrt{3}}},\left(-\frac{p_y}{\sqrt{3}}+D\right)e^{\frac{2}{\sqrt{3}}y}\right]
\end{equation}
where $\alpha$ is an overall scale factor and $x,y$ are the Misner anisotropy parameters with conjugates $D,p_x$ and $p_y$ respectively. In this choice of variables the Bianchi I Hamiltonian becomes 
\begin{equation}
    H_{BI} = \alpha^{-\frac{3}{2}}\left[\frac{1}{2}\left(p_x^2 + p_y^2\right) - \frac{3}{2}D^2\right]
\end{equation}
To this Hamiltonian into the form that is conventionally used in the literature we make a final change of variables
\begin{equation}
    \alpha = \nu^{\frac{2}{3}}, \quad D = \frac{\nu\tau}{2}
\end{equation}
where $\nu$ is the volume factor and it's conjugate momentum $\tau$ is the known as the York time. In these variables, the Bianchi I Hamiltonian is 
\begin{equation}
    H_{BI} = \nu^{-1}\left[-\frac{3}{8}\nu^2\tau^2 + \frac{1}{2}\left(p_x^2 + p_y^2\right)\right]
\end{equation}

The Bianchi IX cosmology is not as straightforward, as in this case the structure constants are given by the Levi-Civita symbol $C^c_{ab} = \epsilon^c_{ab}$. A choice of the invariant 1-form basis in terms of the coordinate basis is given by
\begin{equation}
    \begin{aligned}
        \omega^1 &= -\sin r \, d\theta + \cos r\sin\theta \, d\phi \\
        \omega^2 &= -\cos r \, d\theta - \sin r \sin\theta\ \, d\phi \\
        \omega^3 &= dr + \cos\theta \, d\phi
    \end{aligned}
\end{equation}

Where $(r,\theta,\phi)$ are spherical coordinates. The Hamiltonian and diffeomorphism constraints from eq \ref{HN} and \ref{DN} (which are written in the coordinate basis) become 
\begin{equation}
    \mathbb{H}[N] = \int_{\Sigma_t}d^3x \frac{N\sin\theta}{\sqrt{\gamma}}\left[\text{Tr}\gamma^2 - \frac{1}{2}\left(\text{Tr}\gamma\right)^2 + \pi^{ab}\pi_{ab}-\frac{1}{2}\pi^2\right]
\end{equation}
\begin{equation}
    \mathbb{D}[N^a] = -2\int_{\Sigma_t}d^3x N^a\sin\theta\epsilon^c_{ab}\pi^{bc}\gamma_{bc}
\end{equation}

Just as in the Bianchi I case, the spatially integrated lapse function can be packaged into a variable $n$
\begin{equation}
    n = \int_{\Sigma_t}d^3x \, N\sin\theta
\end{equation}
which enters as a Lagrange multiplier, and thus the overall factor of $n$ can be set to unity in both the Hamiltonian and diffeomorphism constraint giving
\begin{equation}
    \mathbb{H} = \frac{1}{\sqrt{\gamma}}\left(\text{Tr}\gamma^2 - \frac{1}{2}\left(\text{Tr}\gamma\right)^2 + \pi^{ab}\pi_{ab}-\frac{1}{2}\pi^2\right) \approx 0
\end{equation}
\begin{equation}
    \mathbb{N} = -2\epsilon^c_{ab}\pi^{bc}\gamma_{bc} \approx 0
\end{equation}
The diffeomorphism constraint can be written in terms of a commutator that strongly suggests the same gauge fixing should as in the Bianchi I case should be used here
\begin{equation}
    \left[\pi^{ac},\gamma_{cb}\right] \approx 0
\end{equation}
To this end we impose the Taub gauge where the 3-metric and conjugate momentum are diagonal and choose the same Misner anisotropy parameter variables as in eq's \ref{gamma} and \ref{pi}. Following the same transformation of variables one arrives at the Bianchi IX ADM Hamiltonian

\begin{equation}
    H_{BIX} = \nu^{-1}\left[-\frac{3}{8}\nu^2\tau^2 + \frac{1}{2}(p_x^2 + p_y^2)  + \nu^{\frac{4}{3}}V_s(x,y)\right] 
\end{equation}
where the function $V_s(x,y)$ is the ``shape potential" given by
\begin{equation}
   \begin{aligned}
        &V_s(x,y) = f(-\sqrt{3}x + y) + f(\sqrt{3}x+y) + f(-2y) \\
        &f(z) = \frac{1}{2}e^{\sqrt{\frac{2}{3}}z} - e^{-\frac{z}{\sqrt{6}}}
    \end{aligned}
\end{equation}
The shape potential originates from the 3-Ricci scalar on the spatial hypersurface, which in the chosen coordinates is simply the volume factor multiplied by a ``scale-free" part $^3R(\gamma) = \nu^{-\frac{2}{3}}V_s(x,y)$. There is an important comparison between the Bianchi IX and Bianchi I Hamiltonian. One may view the Bianchi IX Hamiltonian as determining the dynamical evolution of a massless particle in 2-dimensions $(x,y)$ under a potential $V_s(x,y)$. The Bianchi I cosmology can be seen as the potential-free case of a Bianchi IX evolution (or vice-versa with Bianchi IX being viewed as Bianchi I evolution under a particle potential).

\subsection{FLRW}
\label{FLRWsect}
In section 2.1 we established that it is possible to reduce the standard symplectic description of a system to a contact one when there exists a scaling symmetry $\mathbf{D}$ such that the Lie drag of the Lagrangian moves between indistinguishable solutions $L_{\mathbf{D}_\Lambda}\mathcal{L} = \Lambda \mathcal{L}$. In this section we will describe the process of contact-reducing the flat ($k=0$) FLRW (+ free massless scalar field) Lagrangian by removing a redundant, non-physical degree of freedom, the volume factor $\nu$ leaving a dynamical system which is described in therms of the physical relational quantities such as the Hubble factor $\nu'/\nu$. We start by considering the action of a scalar field minimally coupled to the FLRW metric
\begin{equation}
    S = \int_\mathbb{R}\int_{\Sigma_t} d^4x \sqrt{-g}\left(R + \frac{1}{2}\dot{\phi}^2 - V(\phi)\right)
\end{equation}

where the spacetime $\mathcal{M} = \mathbb{R} \times \Sigma_t$ is foliated by spacelike hypersurfaces $\Sigma_t$. The metric  $g_{\mu\nu}$ is the FLRW metric with line element $ds^2 = -dt^2 + a(t)^2(dx^2+dy^2+dz^2)$. This is a homogeneous and isotropic universe. The condition of isotropy will later be dropped to study Bianchi cosmologies. 

The Ricci scalar associated with the FLRW metric is 
\begin{equation}
    R = 6\left[\frac{\ddot{a}}{a} + \left(\frac{\dot{a}}{a}\right)^2 - V(\phi)\right]
\end{equation}
In terms of the volume factor $\nu(t) = a(t)^3$ the action is 
\begin{equation}
    S = \int d^4x \hspace{0.25em} \left(-\frac{2}{3}\frac{\dot{\nu}^2}{\nu} +  \frac{1}{2}\nu\dot{\phi}^2 - V(\phi)\right)
\end{equation}
As is customary, we have set to zero the boundary term due to $\ddot{\nu}$. The Lagrangian 
\begin{equation}
\label{L_FLRW}
    \mathcal{L} = \nu\left[-\frac{2}{3}\left(\frac{\dot{\nu}}{\nu}\right)^2 + \frac{1}{2}\dot{\phi}^2 -  V(\phi)\right]
\end{equation}
exhibits a scaling symmetry of the volume factor 
\begin{equation}
\nu = \lambda \bar{\nu} \implies \mathcal{L} = \lambda\bar{\nu}\left[-\frac{2}{3}\left(\frac{\dot{\bar{\nu}}}{\bar{\nu}}\right) + \frac{1}{2}\dot{\phi}^2-V(\phi)\right] = \lambda\bar{\mathcal{L}}, \quad \lambda\in\mathbb{R}/\{0\}
\end{equation}

The CSSS that generates this scaling symmetry is $\bar{\mathbf{D}} = \nu\partial_\nu$ with tangent lift $\mathbf{D} = \nu\partial_\nu + \dot{\nu}\partial_{\dot{\nu}}$
By acting on the Lagrangian with $\mathbf{D}$ one sees clearly that this is already a scaling symmetry of degree 1 (therefore no coordinate transformations are required before forming the Herglotz Lagrangian). As per section 2.1, there exits a Herglotz Lagrangian $\mathcal{L}^H$ given by eq \ref{Herglotz_L} which describes the same physics as $\mathcal{L}$ on the symmetry-reduced contact manifold $TQ/\mathbf{D}$. In the following sections we will switch notation from $S\rightarrow h$ to describe the global coordinate on the contact manifold.

\begin{equation}
    \label{LH_FLRW}
    \begin{aligned}
        &h = \frac{1}{\nu}i_{\mathbf{D}}\mu_L \\
        &\mathcal{L}^H(q,\dot{q},h) = \frac{\partial \mathcal{L}}{\partial \nu} = \frac{3}{8}h^2 + \frac{1}{2}\dot{\phi}^2 - V(\phi)
    \end{aligned}
\end{equation}
The contact Hamiltonian given by the Legendre transformation and contact form are
\begin{equation}
    \label{FLRW_Hc}
    \mathcal{H}^c = -\frac{3}{8}h^2 + \frac{1}{2}p_\phi^2 + V(\phi), \quad \eta = -dh + p_\phi d\phi 
\end{equation}
In the contact formalism, scalar field has been decoupled from the redundant scale factor variable in the symplectic system, however this still does not make $(h,\phi)$ the correct choice of coordinates to describe the manifold geometry near the Big Bang singularity. These variables still diverge as we approach the Big Bang. We will show in section \ref{shapesect} that making a compactification onto shape space provides a suitable description in which the dynamical variables, contact form and Hamiltonian are well defined through the Big Bang.
The Herglotz equations of motion for the Lagrangian \ref{LH_FLRW} are 
\begin{equation}
    \label{FLRW_LH_Phi}
    \frac{d}{dt}\left(\frac{\partial \mathcal{L}^H}{\partial \dot{\phi}}\right) - \frac{\partial \mathcal{L}^H}{\partial \phi} = \frac{\partial\mathcal{L}^H}{\partial h}\frac{\partial\mathcal{L}^H}{\partial\dot{\phi}} \implies \ddot{\phi} +\partial_\phi V = \frac{3}{4}h\dot{\phi}
\end{equation}
\begin{equation}
    \label{FLRW_LH_h}
    \dot{h} = \mathcal{L}^H \implies \dot{h} = \frac{3}{8}h^2 + \frac{1}{2}\dot{\phi}^2 - V(\phi)
\end{equation}
In this example it's quite clear how a contact manifold is a useful description of frictional systems. The equation of motion for $\phi(t)$ \ref{FLRW_LH_Phi} can be seen as a driven harmonic oscillator with a time dependant frictional term $h(t)$. In this case the oscillator is being driven by its potential $V(\phi)$ with friction generated by the Hubble factor $h(t)$.

In the potential-free case these equations can be solved exactly, with solution
\begin{equation}
    h(t) = -\frac{4}{3t}, \quad \phi(t) = \phi_0 + \frac{2}{\sqrt{3}}\ln|t|
\end{equation}
It is explicitly clear that there still exits a divergence of the dynamical variables as we reach the big bang at $t\rightarrow 0$. In the shape space representation, the dynamical variables of the contact Hamiltonian/Herglotz Lagrangian will remain well defined.

\subsection{Bianchi}
\label{Bianchisect}
\subsubsection{Vacuum Bianchi}
The ADM Hamiltonian of a type A Bianchi cosmology with shape potential $V_s(x,y)$ is given by \cite{Jha:2022svf,Corichi:1991qqo,PhysRevLett.22.1071}
\begin{equation}
    \mathcal{H}_B = \nu^{-1}\left[-\frac{3}{8}\nu^2\tau^2 +\frac{1}{2} \left(p_x^2 + p_y^2\right)  + \nu^{\frac{4}{3}}V_s(x,y)\right] 
\end{equation}
where $(x,y)$ are the anisotropy parameters with conjugate momenta $k_i$, $\nu$ is the scale factor with conjugate $\tau$ (York time). For Bianchi I the shape potential is everywhere zero and for Bianchi IX it is given by
\begin{equation}
   \begin{aligned}
        &V_s(x,y) = f(-\sqrt{3}x + y) + f(\sqrt{3}x+y) + f(-2y) \\
        &f(z) = \frac{1}{2}e^{\sqrt{\frac{2}{3}}z} - e^{-\frac{z}{\sqrt{6}}}
    \end{aligned}
\end{equation}
whilst Bianchi cosmologies are most easily studied in the Hamiltonian/ADM formalism, for the purpose of applying the contact-reduction scheme outlined in section 2.1 we will use Lagrangian given by the Legendre transformation
\begin{equation}
    \label{L_Bianchi}
    \mathcal{L} = \nu \left[-\frac{2}{3}\left(\frac{\dot{\nu}}{\nu}\right)^2 + \frac{1}{2}\left(\dot{x}^2 + \dot{y}^2\right) - \nu^{-\frac{2}{3}}V_s(x,y)\right]
\end{equation}
In the Bianchi I case $V_s = 0$ the Lagrangian reduces to 
\begin{equation}
    \mathcal{L} = \nu \left[-\frac{2}{3}\left(\frac{\dot{\nu}}{\nu}\right)^2 + \frac{1}{2}\left(\dot{x}^2 + \dot{y}^2\right)\right]
\end{equation}
which is identical in form to the FLRW Lagrangian except we have the anisotropy parameters as our dynamical variables rather than a scalar field. Contact-reducing this Lagrangian thus follows the exact same steps as in part 2.2. For this section we will show the contact-reduction with the general non-zero shape potential. starting with the Bianchi Lagrangian eq \ref{L_Bianchi}, we see that there is a scaling symmetry $\nu=\lambda\bar{\nu}$, $t = \lambda^\beta\bar{t}$ under which the Lagrangian transforms as
\begin{equation}
    \mathcal{L} = \lambda\bar{\nu}\left[-\frac{2}{3}\lambda^{-2\beta}\left(\frac{\bar{\nu}'}{\bar{\nu}}\right)^2 + \frac{1}{2}\lambda^{-2\beta}(x'^2 + y'^2) - \lambda^{-\frac{2}{3}}\bar{\nu}V_s(x,y)\right], \quad q' = \frac{dq}{d\bar{t}}
\end{equation}
We require $\beta = 1/3$ in order for the shape potential term to scale the same with $\lambda$ as the other two terms in the brackets. Making the CSSS vector field 
\begin{equation}
    \bar{\mathbf{D}} = \nu\frac{\partial}{\partial \nu} + \frac{1}{3}t\frac{\partial}{\partial t}
\end{equation}
The CSSS is only defined up to a non-zero constant factor, so with foresight we will choose to work with 
\begin{equation}
    \bar{\mathbf{D}} = \frac{3}{2}\nu\frac{\partial}{\partial \nu} + \frac{1}{2}t\frac{\partial}{\partial t}
\end{equation}
which has tangent lift 
\begin{equation}
    \mathbf{D} = \frac{3}{2}\nu\frac{\partial }{\partial \nu} + \dot{\nu}\frac{\partial}{\partial\dot{\nu}} - \frac{1}{2}\dot{x}\frac{\partial}{\partial\dot{x}} - \frac{1}{2}\dot{y}\frac{\partial}{\partial\dot{y}}+\frac{1}{2}t\frac{\partial}{\partial t}
\end{equation}
This is a non-isochronal scaling symmetry with degree $1/2$, therefore we require a coordinate transformation and reparameterisation of the time variable in order to form an isochornal, degree 1 scaling symmetry.
Consider the transformation $\rho = \nu^{\frac{2}{3}}$, $d\tau = \rho^{-\frac{1}{2}}dt$ under which the action becomes 
\begin{equation}
    S = \int \mathcal{L}(\nu,\dot{\nu},\mathbf{x},\dot{\mathbf{x}})dt = \int \rho \left[-\frac{3}{2}\left(\frac{\rho'}{\rho}\right)^2 + \frac{1}{2}(x'^2 + y'^2) - V_s(x,y)\right]d\tau, \quad q' = \frac{dq}{d\tau}
\end{equation}
We identify the transformed Lagrangian
\begin{equation}
    \mathcal{L} = \rho \left[-\frac{3}{2}\left(\frac{\rho'}{\rho}\right)^2 + \frac{1}{2}(x'^2 + y'^2) - V_s(x,y)\right]
\end{equation}
which clearly has a scaling symmetry $\mathbf{D} = \rho\partial_\rho + \rho'\partial_{\rho'}$ of degree 1. Thus there is a Herglotz Lagrangian $\mathcal{L}^H$ on $TQ/\mathbf{D}$ given by
\begin{equation}
    \begin{aligned}
        &h =\frac{1}{\rho} i_{\mathbf{D}}\mu_L = \frac{\partial \mathcal{L}}{\partial\rho'} = -3\frac{\rho'}{\rho} \\
        &\mathcal{L}^H(\mathbf{x},\dot{\mathbf{x}},h) = \frac{\partial\mathcal{L}}{\partial \rho} = \frac{1}{6}h^2 + \frac{1}{2}(x'^2+y'^2)- V_s(x,y)
    \end{aligned}
\end{equation}
In this contact manifold description, the volume factor of the universe becomes decoupled in the Lagrangian from the anisotropy parameters and shape potential. Herglotz equations of motion for this system are similar in form to those of the FLRW with scalar field in section 2.2, but with a shape potential of the anisotropy parameters rather than a scalar field potential.
The Kasner universe is the vacuum Bianchi I cosmology, which corresponds to the potential-free case $V_s = 0$ for which the equations of motion reduce to (we now use the ordinary $\dot{q}$ to denote time derivatives rather than $q'$) 
\begin{equation}
    \begin{aligned}
        &\dot{h} = \frac{1}{6}h^2 + \frac{1}{2}\left(\dot{x}^2 + \dot{y}^2\right) \\
        &\ddot{q} = \frac{1}{3}h\dot{q}, \quad q = x,y
    \end{aligned}
\end{equation}
These equations can be solved analytically, with solutions
\begin{equation}
    h(t) = -\frac{3}{t}
\end{equation}
\begin{equation}
    q(t) = q_0 + C_q\ln|t|, \quad \sum_q C_q^2 = 3
\end{equation}

At this point we note that, whilst the quantities $h(t),x(t),y(t)$ may diverge as $t\rightarrow 0$, relational quantities like $x(t)/y(t)$ remain finite, as well as 
\begin{equation}
    h(t)e^{-\frac{1}{2}\sum_q\frac{q(t)}{C_q}}
\end{equation}
This will motivate our choice of variables when we project the system onto shape space in section 3.

Once the Herglotz Lagrangian has been obtained it is possible to move to the contact Hamiltonian representaion by a Legendre transformation
\begin{equation}
    \mathcal{H}^c = -\frac{1}{6}h^2 + \frac{1}{2}\left(p_x^2 + p_y^2\right) + V_s(x,y)
\end{equation} 
with contact form 
\begin{equation}
    \eta = -dh + p_xdx + p_ydy
\end{equation}
The Hamiltonian representation will be used in making the projection to shape space in section \ref{shapesect}.

\subsubsection{Minimally Coupled Scalar Field}
The case of a scalar field minimally coupled to a Bianchi cosmology changes the procedure slightly, since the scalar field potential $V(\phi)$ couples to the scale factor differently to the shape potential $V_s(x,y)$
\begin{equation}
    \label{L_Bianchi_Scalar}
    \mathcal{L} = \nu \left[-\frac{2}{3}\left(\frac{\dot{\nu}}{\nu}\right)^2  + \frac{1}{2}\dot{\phi}^2+ \frac{1}{2}\left(\dot{x}^2 + \dot{y}^2\right) - \nu^{-\frac{2}{3}}V_s(x,y) - V(\phi)\right]
\end{equation}

In the case of Bianchi I, the shape potential is everywhere zero and thus the Lagrangian reduces to a form that immediately has a scaling symmetry of degree 1 associated with the scale factor $\nu$. Therefore the exact same procedure may be followed in section 2.2, whereby the contact Hamiltonian for Bianchi I + matter is given by 
\begin{equation}
\label{HBI}
    \mathcal{H}^c_{BI} = -\frac{1}{6}h^2 + \frac{1}{2}(p_x^2 + p_y^2) + \frac{1}{2}p_\phi^2 + V(\phi)
\end{equation}

For Bianchi cosmologies with a shape potential that is not everywhere zero, the scaling symmetry we previously identified in section 2.2.1 $\nu = \lambda\bar{\nu}$, $t = \lambda^{\frac{1}{2}}\bar{t}$ is not a symmetry for this Lagrangian in eq \ref{L_Bianchi_Scalar}. However, the scaling symmetry can be restored by introducing a velocity term $\dot{k}$ and forming a new Lagrangian 
\begin{equation}
    \label{L_Bianchi_New}
    \mathcal{L}_* = \nu \left[-\frac{2}{3}\left(\frac{\dot{\nu}}{\nu}\right)^2  + \frac{1}{2}\dot{\phi}^2+ \frac{1}{2}\left(\dot{x}^2 + \dot{y}^2\right) - \nu^{-\frac{2}{3}}V_s(x,y)\right] - \sqrt{\nu\dot{k}V(\phi)}
\end{equation}
The new variable $k$ is a cyclic coordinate with constant momentum, it is possible to choose an appropriate boundary condition for $\dot{k}$ such that the equations of motion generated by $\mathcal{L}$ coincide with those generated by the new Lagrangian $\mathcal{L}_*$. The advantage of working with $\mathcal{L}_*$ is that it has the scaling symmetry we require in order to make a contact-reduction. The equations of motion generated by the original Lagrangian \ref{L_Bianchi_Scalar} are 
\begin{equation}
    \begin{aligned}
        -\frac{4}{3}\frac{d}{dt}\left(\frac{\dot{\nu}}{\nu}\right) &= \frac{2}{3}\left(\frac{\dot{\nu}}{\nu}\right)^2 + \frac{1}{2}\left(\dot{x}^2 + \dot{y}^2\right) + \frac{1}{2}\dot{\phi}^2 - \frac{1}{3}\nu^{\frac{1}{3}}V_s - V(\phi) \\
        \frac{d}{dt}\left(\nu\dot{q}\right) &= -\nu^{\frac{1}{3}}\partial_q V_s, \quad q = x,y \\
        \frac{d}{dt}\left(\nu\dot{\phi}\right) &= -\nu\partial_\phi V(\phi)
    \end{aligned}
\end{equation}
and the equations of motion generated by the new Lagrangian $\mathcal{L}_*$ are 
\begin{equation}
    \begin{aligned}
        -\frac{4}{3}\frac{d}{dt}\left(\frac{\dot{\nu}}{\nu}\right) &= \frac{2}{3}\left(\frac{\dot{\nu}}{\nu}\right)^2 + \frac{1}{2}\left(\dot{x}^2 + \dot{y}^2\right) + \frac{1}{2}\dot{\phi}^2 - \frac{1}{3}\nu^{\frac{1}{3}}V_s \\
        \frac{d}{dt}\left(\nu\dot{q}\right) &= -\nu^{\frac{1}{3}}\partial_q V_s, \quad q = x,y \\
        \frac{d}{dt}\left(\nu\dot{\phi}\right) &= -\frac{1}{2}\sqrt{\frac{\nu\dot{k}}{V(\phi)}}\partial_\phi V(\phi) \\
        C &= -\frac{1}{2}\sqrt{\frac{\nu V(\phi)}{\dot{k}}}, \quad C = \text{constant}
    \end{aligned}
\end{equation}
One can see that choosing the boundary condition $C = -\frac{1}{4}$ restores the original equations of motion. Choosing a boundary condition on $4\dot{k} = C^{-2}V(\phi)\nu$ is equivalent to choosing an overall scale of the potential field potential $V(\phi)$, which would have to be chosen in order to specify a solution. So while it may appear at first that the new Lagrangian $\mathcal{L}_*$ requires additional information to specify a solution compared to $\mathcal{L}$, in in fact does not, only the choice in scale of the potential is made explicit through $\dot{k}$, rather than implicit in the definition of $V(\phi)$. The new Lagrangian $\mathcal{L}_*$ now generates the same dynamics as the original in eq \ref{L_Bianchi_Scalar} whilst retaining the desired scaling symmetry $\nu = \lambda\bar{\nu}$, $t = \lambda^\frac{1}{3}\bar{t}$. Choosing to work with the CSSS
\begin{equation}
    \bar{\mathbf{D}} = \frac{3}{2}\nu\frac{\partial }{\partial \nu} + \frac{1}{2}t\frac{\partial}{\partial t}
\end{equation}
we have the tangent lift 
\begin{equation}
    \mathbf{D} = \frac{3}{2}\nu\frac{\partial }{\partial \nu} + \dot{\nu}\frac{\partial}{\partial\dot{\nu}} - \frac{1}{2}\dot{x}\frac{\partial}{\partial\dot{x}} - \frac{1}{2}\dot{y}\frac{\partial}{\partial\dot{y}} -\frac{1}{2}\dot{\phi}\frac{\partial}{\partial\dot{\phi}} - \frac{1}{2}\dot{k}\frac{\partial}{\partial \dot{k}} +\frac{1}{2}t\frac{\partial}{\partial t}
\end{equation}
We again make a coordinate transformation and reparameterisation of the time coordinate under
\begin{equation}
    \rho = \nu^{\frac{2}{3}}, \quad dt = \rho^{-\frac{1}{2}}d\tau
\end{equation}

which the action transforms as 
\begin{equation}
    S = \int \mathcal{L}_* dt = \int \rho\left[-\frac{3}{2}\left(\frac{\rho'}{\rho}\right)^2 + \frac{1}{2}\left(x'^2 + y'^2\right) + \frac{1}{2}\phi'^2 - V_s(x,y) - \sqrt{k'V(\phi)}\right]d\tau
\end{equation}
and identify the transformed Lagrangian as 
\begin{equation}
    \mathcal{L} = \rho\left[-\frac{3}{2}\left(\frac{\rho'}{\rho}\right)^2 + \frac{1}{2}\left(x'^2 + y'^2\right) + \frac{1}{2}\phi'^2 - V_s(x,y) - \sqrt{k'V(\phi)}\right]
\end{equation}
This has scaling symmetry $\mathbf{D} = \rho\partial_\rho + \rho'\partial_{\rho'}$ of degree 1, hence there is a Herglotz Lagrangian given by
\begin{equation}
    \begin{aligned}
        &h = \frac{1}{\rho}i_\mathbf{D}\mu_L = -3\frac{\rho'}{\rho} \\
        & \mathcal{L}^H = \frac{\partial \mathcal{L}}{\partial\rho} = \frac{1}{6}h^2 + \frac{1}{2}\left(x'^2 + y'^2\right) + \frac{1}{2}\phi'^2 - V_s(x,y) - \sqrt{k'V(\phi)}
    \end{aligned}
\end{equation}
The contact Hamiltonian obtained from a Legendre transformation and contact form are 
\begin{equation}
    \begin{aligned}
        \mathcal{H}^c &= -\frac{1}{6}h^2 + \frac{1}{2}\left(p_x^2 + p_y^2\right) + \frac{1}{2}p_\phi^2 + V_s(x,y) - \frac{V(\phi)}{4p_k} \\
        \eta &= -dh + p_kdk + p_\phi d\phi + p_xdx + p_ydy
    \end{aligned}
\end{equation}
The cast the Hamiltonian into the form that will be used for the shape space projection, we make one last parity transformation of the $p_k$ momentum and absorb the factor constant factor $4p_k \rightarrow -p_k$ so that we may work with a positive $p_k$. Thus we arrive at 
\begin{equation}
\label{Contact_Bianchi_Scalar_H}
    \begin{aligned}
        \mathcal{H}^c &= -\frac{1}{6}h^2 + \frac{1}{2}\left(p_x^2 + p_y^2\right) + \frac{1}{2}p_\phi^2 + V_s(x,y) + \frac{V(\phi)}{p_k} \\
        \eta &= -dh -\frac{1}{4}p_kdk + p_\phi d\phi + p_xdx + p_ydy
    \end{aligned}
\end{equation}
We now have contact Hamiltonians for FLRW + scalar field, vacuum Bianchi and Bianchi + scalar field cosmologies. Although the overall scale factor of the universe $\nu(t)$ is decoupled from the dynamical variables in the contact description, as demonstrated through the cases of potential-free FLRW and Kasner solutions, the dynamical variables still diverge at the initial singularity. There is no apriori reason why this choice of coordinates on the contact manifold should be a suitable one for describing the universe at the Big Bang. In section 3 we will outline a procedure that projects the system onto shape space. We will then prove that the equations of motion have unique solutions in a neighbourhood of the initial singularity via the Picard-Lindelöf theorem.

\clearpage

\section{Shape Space Projection and Proof of Existence and Uniqueness Through $\beta = \pi/2$}
\label{shapesect}
In the following section we will take the symmetry-reduced models of FLRW and Bianchi I and IX on contact manifolds, and describe the procedure for projecting them onto shape space. The configuration space coordinates of the symplectic systems that one starts with have as their spatial manifolds, $\Sigma$ n-dimensional real space $\mathbb{R}^n$. In forming the contact system, this space is quotiented by a scaling symmetry resulting in n-dimensional real-projective space $\mathbb{RP}^n$. An important property of the gnomonic projection is that it maps straight lines in the plane to great circles on the sphere. Thus we see the geodesic motion of a free particle in the plane represented by a great circle in shape space coordinates. For a concrete example of this, we refer the reader to appendix \ref{Gno}.

The projection onto shape space of the contact Hamiltonian systems under consideration is a gnomonic projection of the dynamical variables in $\mathbb{RP}^n$ onto the unit sphere $S^n$. The dimensionality $n$ will depend on the number of dynamical variables. The gnomonic projection constitutes of a compactification of $\mathbb{RP}^n$ onto an $n$ sphere. In choosing the sign of the triads used to describe the geometry, we implicitly decide on an orientation of the manifold, which does not affect the physical dynamics. The gnomonic projection maps points on the surface of an n-sphere to a tangent plane at one of the poles by drawing a straight line from the centre of the sphere through the surface point and intersects it with the plane. We must form a double cover in order to describe both orientations of the real projective space. In $n=2$ dimensions, this consists of two distinctly oriented tangent planes at antipodal points of an $S^2$ surface. The gnomonic projection maps the asymptotic boundaries of the planes to the equator of the sphere, forming a border between the two planes.

We first consider the cases of FLRW with one and then two scalar fields and show how the compactification to shape space works in the simplest cases. Once the Hamiltonian has been written in shape space variables, we show that there exist unique, smooth solutions to the equations of motion at the big bang, which is mapped to the equator of $S^n$ under the gnomonic projection. We then progress to the Bianchi I and IX cosmologies at show the same result.

\subsection{FLRW + 1 scalar field}
We start with the contact Hamiltonian and contact form derived in eq \ref{FLRW_Hc} for FLRW + a minimally coupled scalar field
\begin{equation}
    \begin{aligned}
        \mathcal{H}^c &= -\frac{3}{8}h^2 + \frac{1}{2}p_\phi^2 + V(\phi) \\
        \eta &= -dh + p_\phi d\phi
    \end{aligned}
\end{equation}
Consider the mapping of $\phi$ on the real line onto $S^1$ through a gnomonic projection
\begin{equation}
    \phi = |\tan\beta|
\end{equation}
Under the gnomonic projection, the initial singularity is mapped to $|\beta| = \pi/2$.
We will also make a transformation of the Hubble factor $h(t)$.
\begin{equation}
    \label{FLRW_h_Transform}
    h = se^{m+a|\tan\beta|}, \quad s = \text{Sign}(\tan\beta), \quad a = \text{constant}
\end{equation}
Although we expect $h(t)$ to diverge as we approach the singularity, the global coordinate $m(t)$ will tend to a finite value. Next we compute the transformed contact form 
\begin{equation}
    \eta = -h\left(dm + sa\sec^2\beta d\beta\right) + sp_\phi\sec^2\beta d\beta
\end{equation}
We may divide both the contact form, and shortly the Hamiltonian by a non-zero factor, as this is equivalent to making a change of lapse.
In particular we choose
\begin{equation}
    \label{FLRW_1_c_Form}
    \begin{aligned}
        \eta\rightarrow \frac{\eta}{sh} &= -sdm + \left(\frac{p_\phi}{h} - a\right)\sec^2\beta d\beta \\
        &= -sdm + \left(s\chi - a\right)\sec^2\beta d\beta
    \end{aligned}
\end{equation}
where we have defined $\chi = p_\phi/|h|$. From the contact form \ref{FLRW_1_c_Form}, the $p_\beta$ momentum associated with $\beta$ can be identified as
\begin{equation}
    p_\beta \left(s\chi - a\right)\sec^2\beta
\end{equation}
In these new variables the contact Hamiltonian is 
\begin{equation}
    \label{FLRW_1_H_a}
    \mathcal{H}^c = \frac{1}{2}h^2\left(p_\beta^2\cos^4\beta + 2ap_\beta\cos^2\beta + a^2 - \frac{3}{4}\right) + V(\phi)
\end{equation}
The constant $a$ may be set freely, a convenient choice of $a = \sqrt{3}/2$ cancels the constant term in eq \ref{FLRW_1_H_a} which diverges like $h^2$. Finally, choosing to divide the contact Hamiltonian by a factor of $h^2\cos^2\beta$ (equivalent to a change of lapse) we arrive at the shape space Hamiltonian for FLRW + a minimally coupled scalar field
\begin{equation}
\label{H_FLRW_1}
    \begin{aligned}
        \mathcal{H} &= \frac{\mathcal{H}^c}{h^2\cos^2\beta} = \frac{1}{2}p_\beta^2\cos^2\beta + \frac{\sqrt{3}}{2}p_\beta + U(\beta)e^{-2m} \\
        U(\beta) &= V(\beta)e^{-\sqrt{3}|\tan\beta|}\sec^2\beta
    \end{aligned}
\end{equation}
Heuristically, we see that provided the potential $V(\beta)$ does not grow faster than the exponential factor $e^{-\sqrt{3}}|\tan\beta|$, any divergence from the potential and the $\sec^2\beta$ factor as $\beta\rightarrow \pi/2$ will be exponentially suppressed, keeping $U(\beta)$ finite at the initial singularity. The contact Hamiltonian equations of motion for this system \ref{Conact_eom} are 
\begin{equation}
    \begin{aligned}
        \dot{\beta} &= p_\beta\cos^2\beta + \frac{\sqrt{3}}{2} \\
        \dot{p}_\beta &= p_\beta^2\cos\beta\sin\beta + \left(2sU p_\beta - \partial_\beta U\right)e^{-2m} \\
        \dot{m} &= -s\left(\frac{\sqrt{3}}{2}p_\beta + 2Ue^{-2m}\right)
    \end{aligned}
\end{equation}
Where in the last line we have used the fact that the ADM Hamiltonian of general relativity satisfies the constraint $\mathcal{H}:=0$ on the spacial hypersurfaces $\Sigma_t$. In the potential free-case this system of ODE's can be solved exactly.
Firstly the Hamiltonian constraint must be satisfied at all times. 
\begin{equation}
    \frac{1}{2}p_\beta^2\left(p_\beta\cos^2\beta + \sqrt{3}\right) = 0
\end{equation}
Thus there is one non-physical diverging solution $p_\beta = -\sqrt{3}\sec^2\beta$ and one that remains finite at the initial singularity $p_\beta = 0$. Although in this case the finite solution is a constant $p_\beta = 0$, we will show shortly that this not need the case, one may have a finite solution with non-zero potential and momentum. The $p_\beta = 0$ solution gives
\begin{equation}
    \begin{aligned}
        \dot{\beta} &= \frac{\sqrt{3}}{2}, \implies \beta(t) = \beta_0 + \frac{\sqrt{3}}{2}t \\
        \dot{m} &= 0 \implies m(t) = m_0
    \end{aligned}
\end{equation}
Thus we reach the initial singularity at $\beta = \pi/2$ in finite proper time. At this point the dynamical variables, Hamiltonian and contact form remain finite and well defined. In the next section 2.2, we will prove in detail that the Picard-Lindelöf theorem is satisfied in a neighbourhood of $\beta = \pi/2$.

\subsection{FLRW + 2 Scalar Fields}
In anticipation of projecting the vacuum Bianchi contact Hamiltonian, which contains two dynamical fields $x(t)$ and $y(t)$, we shall first show how the projection to $S^2$ works with a contact FLRW + 2 scalar fields cosmology. The contact-reduction results of section 2.2 can be arbitrarily extended to $n$ scalar fields in the usual way. The field $\phi$ is simply replaced with a vector of fields $\underline{\phi}$ and $\dot{\phi}^2$ becomes the Euclidean norm squared of $\underline{\dot{\phi}}$. The contact Hamiltonian for FLRW + 2 scalar fields is thus
\begin{equation}
\label{FLRW_2_Hc}
    \mathcal{H}^c = -\frac{3}{8}h^2 + \frac{1}{2}\left(p_1^2 + p_2^2\right) + V(\phi_1,\phi_2)
\end{equation}

In section 3.1 we had a single field and made a gnomonic projection of the real (projective) line onto $S^1$, here there are two fields and thus it required a gnomonic projection of $\mathbb{RP}^2$ to $S^2$ given by
\begin{equation}
    \begin{pmatrix}
        \phi_1 \\
        \phi_2
    \end{pmatrix} = |\tan\beta| 
    \begin{pmatrix}
        \cos\alpha \\
        \sin\alpha
    \end{pmatrix}
\end{equation}
and a projection of the momenta into polar coordinates
\begin{equation}
    \begin{pmatrix}
        p_1 \\
        p_2 
    \end{pmatrix} = p
    \begin{pmatrix}
        \cos\theta \\
        \sin\theta
    \end{pmatrix}
\end{equation}
We make the same transformation of the Hubble factor as in eq \ref{FLRW_h_Transform}, but now with the foresight to fix $a = \sqrt{3}/2$
\begin{equation}
    h = se^{m+\frac{\sqrt{3}}{2}|\tan\beta|}, \quad s = \text{Sign}(\tan\beta)
\end{equation}
The contact form under this transformation becomes
\begin{equation}
    \eta = -h\left(dm + s\frac{\sqrt{3}}{2}\sec^2\beta\right) + sp\cos(\theta-\alpha)\sec^2\beta d\beta + p\sin(\theta-\alpha)|\tan\beta|d\alpha
\end{equation}
again defining the variable ratio $\chi = p/|h|$ and scaling the contact form by the non-zero factor $\frac{1}{sh}$ gives
\begin{equation}
    \eta \rightarrow \frac{\eta}{sh} = -sdm + \left[s\chi\cos(\theta-\alpha)-\frac{\sqrt{3}}{2}\right]\sec^2\beta d\beta + \chi|\tan\beta|\sin(\theta - \alpha)d\alpha
\end{equation}
and identify the momenta 
\begin{equation}
    p_\beta = \left[s\chi\cos(\theta-\alpha)-\frac{\sqrt{3}}{2}\right]\sec^2\beta
\end{equation}
\begin{equation}
    p_\alpha = \chi|\tan\beta|\sin(\theta - \alpha)
\end{equation}
Returning to the contact Hamiltonian, it is written in terms of the new variables as 
\begin{equation}
    \mathcal{H}^c = -\frac{1}{2}h^2\left(\chi^2 - \frac{3}{4}\right) + V(\alpha,\beta)
\end{equation}
Once more, rescaling by the non-zero factor $h^{-2}\sec^2\beta$, we arrive at the shape space Hamiltonian for FLRW + 2 scalar fields
\begin{equation}
\label{FLRW_2_H}
    \begin{aligned}
        &\mathcal{H} = \frac{\mathcal{H}^c}{h^2\cos^2\beta} = \frac{1}{2}p_\beta^2\cos^2\beta + \frac{\sqrt{3}}{2}p_\beta + \frac{p_\alpha^2}{2\sin^2\beta} + U(\alpha,\beta)e^{-2m} \\
        &U(\alpha,\beta) = V(\alpha,\beta)e^{-\sqrt{3}|\tan\beta|}\sec^2\beta
    \end{aligned}
\end{equation}
With equations of motion
\begin{equation}
    \begin{aligned}
        &\dot{\alpha} = \frac{p_\alpha}{\sin^2\beta} \\
        &\dot{\beta} = p_\beta\cos^2\beta + \frac{\sqrt{3}}{2} \\
        &\dot{p}_\alpha = \left(2sUp_\alpha-\partial_\alpha U\right)e^{-2m} \\
        &\dot{p}_\beta = p_\beta^2\cos\beta\sin\beta + \left(2sUp_\beta-\partial_\beta U\right)e^{-2m} \\
        &\dot{m} = -s\left(\frac{\sqrt{3}}{2}p_\beta + 2Ue^{-2m}\right)
    \end{aligned}
\end{equation}
Just as with the single field case in section 3.1, we can analyse the potential-free cosmology analytically. In this case the equations of motion reduce to 
\begin{equation}
    \begin{aligned}
        &\dot{\alpha} = \frac{p_\alpha}{\sin^2\beta} \\
        &\dot{\beta} = p_\beta\cos^2\beta + \frac{\sqrt{3}}{2} \\
        &p_\alpha =  \text{constant} \\
        &\dot{p}_\beta = p_\beta^2\cos\beta\sin\beta  \\
        &\dot{m} = -s\frac{\sqrt{3}}{2}p_\beta
    \end{aligned}
\end{equation}
The Hamiltonian constraint $\mathcal{H}:=0 $ must also be enforced. For the shape space Hamiltonian \ref{FLRW_2_H}, we can think of the constraint as a quadratic equation in $p_\beta$
\begin{equation}
    \sin^2\beta\cos^2\beta p_\beta^2 + \sqrt{3}\sin^2\beta p_\beta + p_\alpha^2 = 0
\end{equation}
with solutions 
\begin{equation}
    p_\beta^\pm = \frac{\sqrt{3}}{2}\sec^2\beta\left(-1\pm \sqrt{1 - \frac{4}{3}p_\alpha^2\cot^2\beta}\right)
\end{equation}
At a first glance it may seem that both solutions become undefined at $\beta = \pi/2$ due to the factor of $\sec^2\beta$, but if one considers the series expansion of the $p_\beta^+$ solutions, valid for $\tan^2\beta \geq 4p_\alpha^2/3$
\begin{equation}
    \begin{aligned}
        \label{FLRW_2_PB}
        p_\beta^+ &= \frac{\sqrt{3}}{2}\sec^2\beta \left[-1+ \left(1 - 2p_\alpha^2\cot^2\beta - \frac{1}{2}\cot^4\beta + \mathcal{O}(\cot^6\beta)\right)\right] \\
        &= -\frac{p_\alpha^2}{2\sqrt{3}\sin^2\beta}\left[2 + \frac{1}{2}p_\alpha^2\cot^2\beta + \mathcal{O}(\cot^4\beta)\right]
    \end{aligned}
\end{equation}
In the series expansion it is clear that $p_\beta^+$ is finite through the initial singularity at $\beta = \pi/2$. Just as $p_\beta$ has been parameterised in terms of $\beta$ in eq \ref{FLRW_2_PB}, we may also look for the solution $\alpha(\beta)$.
\begin{equation}
    \label{FLRW_2_DAB}
    \frac{d\alpha}{d\beta} = \frac{\dot{\alpha}}{\dot{\beta}} = \frac{2p_\alpha}{\sin^2\beta\sqrt{3-4p_\alpha^2\cot^2\beta}}
\end{equation}
Equation \ref{FLRW_2_DAB} has the solution
\begin{equation}
\label{FLRW_2_AB}
    \frac{2p_\alpha}{\sqrt{3}}\cot\beta = \sin[\sqrt{2}(\alpha_0-\alpha)]
\end{equation}
Thus we have a solution of dynamical variables, Hamiltonian and contact form that remain well defined through the initial singularity. Furthermore, equation \ref{FLRW_2_AB} is the equation of a great circle on shape space. This is to be expected since the potential-free motion corresponds to a straight line in the $(\phi_1,\phi_2)$ plane and gnomonic projections map straight lines in the plane to great circles on the sphere.

\subsection{Bianchi I}

\subsection{Vacuum Bianchi}
\label{VB}
We know turn to the case of a contact-reduced vacuum Bianchi spacetime with shape potential $V_s(x,y)$. As shown in section 2.3, the contact Hamiltonian is given by 
\begin{equation}
    \mathcal{H}^c = -\frac{1}{6}h^2 + \frac{1}{2}\left(\dot{x}^2 + \dot{y}^2\right) + V_s(x,y)
\end{equation}
By simply inspecting the Hamiltonian, it is clearly functionally the same as the FLRW + 2 scalar field Hamiltonian in eq. \ref{FLRW_2_Hc}. The projection to shape space thus follows the exact same procedure as that described in section 3.2, setting $a = 1/\sqrt{3}$ to account for the coefficient of $-\frac{1}{6}$ on the Hubble factor. The shape space Hamiltonian for a vacuum Bianchi cosmology is thus 
\begin{equation}
    \label{Bianchi_H}
    \begin{aligned}
        &\mathcal{H} = \frac{1}{2}p_\beta^2\cos^2\beta + \frac{1}{\sqrt{3}}p_\beta + \frac{p_\alpha^2}{2\sin^2\beta} + U(\alpha,\beta)e^{-2m} \\
        &U(\alpha,\beta) = V(\alpha,\beta)e^{-\frac{2}{\sqrt{3}}|\tan\beta|}\sec^2\beta
    \end{aligned} 
\end{equation}
with equations of motion
\begin{equation}
    \label{Bianchi_Vacuum_eom}
    \begin{aligned}
        &\dot{\alpha} = \frac{p_\alpha}{\sin^2\beta} \\
        &\dot{\beta} = p_\beta\cos^2\beta + \frac{1}{\sqrt{3}} \\
        &\dot{p}_\alpha = \left(2sUp_\alpha-\partial_\alpha U\right)e^{-2m} \\
        &\dot{p}_\beta = p_\beta^2\cos\beta\sin\beta + \left(2sUp_\beta-\partial_\beta U\right)e^{-2m} \\
        &\dot{m} = -s\left(\frac{1}{\sqrt{3}}p_\beta + 2Ue^{-2m}\right)
    \end{aligned}
\end{equation}
The Kasner solutions thus correspond to great circles on $S^2$ given by
\begin{equation}
    \label{Kasner_GC}
    \sqrt{3}p_\alpha\cot\beta = \sin[2(\alpha_0-\alpha)]
\end{equation}
The Picard-Lindelöf theorem states that, given an ordinary differential equation of the form
\begin{equation}
    r'(t) = f(t,r), \quad r(t_0) = r_0
\end{equation}
there exists a unique, local solution if $f(t,r)$ is continuous in $t$ and locally Lipschitz continuous in $r$. This extends to a system of ODE's 
\begin{equation}
    r_i'(t) = f_i(t,\mathbf{r}), \quad r_i(t_0) = r_{i0}
\end{equation}
where the functions $f_i(t,\mathbf{r})$ are required to be continuous in $r$ and locally Lipschitz continuous in $r_i$.
The Picard-Lindelöf theorem applies immediately to the equations of motion \ref{Bianchi_Vacuum_eom}, the RHS's are not explicitly dependant on time and are locally Lipschitz around $\beta = \pi/2$ provided that the potential $U(\alpha,\beta)$ are its derivatives are locally bounded around $\beta = \pi/2$. Assuming that the potential satisfies these conditions, there exists a unique local solution of the equations of motion that continues smoothly through the big bang. In the case of the vaccum Bianchi I cosmology, this is satisfied trivially as the shape potential is everywhere zero. However for Bianchi IX the shape potential is such that the exponential suppression in $U(\alpha,\beta)$ is killed off exactly. The shape potential is given by
\begin{equation}
    V_s(x,y) = \frac{1}{2}e^{\frac{2}{\sqrt{3}}y}\left(e^{2x}+e^{-2x}\right) - e^{-\frac{1}{\sqrt{3}}y}\left(e^x + e^{-x}\right) + \frac{1}{2}e^{-\frac{4}{\sqrt{3}}y} - e^{\frac{2}{\sqrt{3}}y}
\end{equation}
In terms of the shape space coordinates $(\alpha,\beta)$ this can be written as
\begin{equation}
    V_s(\alpha,\beta) = \frac{1}{2}e^{f_1(\alpha)|\tan\beta|} + \frac{1}{2}e^{f_2(\alpha)|\tan\beta|} - e^{f_3(\alpha)|\tan\beta|} - e^{f_4(\alpha)|\tan\beta|} + \frac{1}{2}e^{f_5(\alpha)|\tan\beta|} - e^{f_6(\alpha)|\tan\beta|}
\end{equation}
The functions $f_i(\alpha)$ are given by
\begin{equation}
\label{f_i}
    \begin{aligned}
        & f_1(\alpha) = \frac{2}{\sqrt{3}}\sin\alpha + 2\cos\alpha, \quad f_2(\alpha) = \frac{2}{\sqrt{3}}\sin\alpha - 2\cos\alpha \\
        & f_3(\alpha) = -\frac{1}{\sqrt{3}}\sin\alpha + \cos\alpha, \quad f_4(\alpha) = -\frac{1}{\sqrt{3}}\sin\alpha - \cos\alpha \\
        & f_5(\alpha) = -\frac{4}{\sqrt{3}}\sin\alpha , \quad f_6(\alpha) = \frac{2}{\sqrt{3}}\sin\alpha
    \end{aligned}
\end{equation}
One can see that clearly, some of these functions will surpass $f_i(\alpha) = 2/\sqrt{3}$, cancelling the exponential suppression of $e^{-\frac{2}{\sqrt{3}}|\tan\beta|}$ in $U(\alpha,\beta)$.

\begin{figure}[h]
    \centering
    \includegraphics[scale = 0.7]{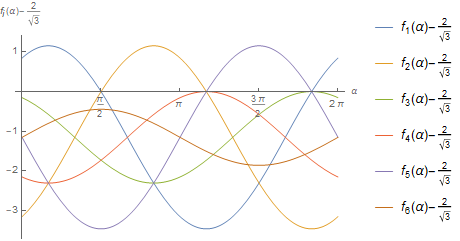}
    \caption{Plots of the functions $f_i(\alpha) - \frac{2}{\sqrt{3}}$ for $0\leq \alpha \leq 2\pi$.}
    \label{F_Plots}
\end{figure}
In figure \ref{F_Plots} we plot the functions $f_i(\alpha)$ over the entire range $0\leq \alpha\leq 2\pi$. Fro all values of $\alpha$, there is at least one functions that is greater than or equal to $2/\sqrt{3}$. So it is not possible in vacuum Bianchi IX to find a subset of $\alpha \in [0,2\pi]$ for which the exponential suppression in $U(\alpha,\beta)$ is not cancelled out.

The fact that local Lipschitz continuity is not in general satisfied for the Vacuum Bianchi IX cosmology should not come as a surprise, as it is well known that Bianchi IX cannot achieve quiescence without the presence of a matter field \cite{Jha:2022svf,PhysRevLett.22.1071,Ashtekar:2011ck}. The spacetime geometry goes through infinitely many Kasner epochs before reaching the singularity. 

\subsection{Bianchi + Scalar Field}

Having seen in section in section 3.3 how Lipschitz continuity fails in the case of Vaccum Bianchi, we are now motivated to consider the shape space projection of the contact Bianchi + scalar field Hamiltonians.

The simplest case is that of Bianchi I + scalar field, since the shape potential is everywhere zero. The contact Hamiltonian for such a system is given by eq. \ref{HBI}
\begin{equation}
    \mathcal{H}^c_{BI} = -\frac{1}{6}h^2 + \frac{1}{2}(p_x^2 + p_y^2) + \frac{1}{2}p_\phi^2 + V(\phi)
\end{equation}
with contact form 
\begin{equation}
    \eta = -dh + p_\phi d\phi + p_xdx + p_ydy
\end{equation}
The configuration space variables $(\phi,x,y)$ are projected onto $S^3$ through the gnomonic projection
\begin{equation}
\label{S3_projection}
    \begin{pmatrix}
        \phi \\
        x \\
        y \\
    \end{pmatrix} = |\tan\beta|
    \begin{pmatrix}
        \cos\gamma \\
        \sin\gamma\cos\alpha \\
        \sin\gamma\sin\alpha
    \end{pmatrix}, \quad 
    \begin{pmatrix}
        p_\phi \\
        p_x \\
        p_y
    \end{pmatrix} = p
    \begin{pmatrix}
        \cos\theta \\
        \sin\theta\cos\psi \\
        \sin\theta\sin\psi
    \end{pmatrix}
\end{equation}
and the Hubble factor is transformed as 
\begin{equation}
     h = se^{m + \frac{1}{\sqrt{3}}|\tan\beta|}, \quad s = \text{Sign}(\tan\beta)
\end{equation}
Under this set of transformations, the rescaled contact form and Hamiltonian become
\begin{equation}
        \eta \rightarrow \frac{\eta}{sh} = -sdm + p_\alpha d\alpha + p_\beta d\beta + p_\gamma d\gamma
\end{equation}
\begin{equation}
\label{shape_H_BI}
    \begin{aligned}
        &\mathcal{H}_{BI} = \frac{\mathcal{H}_{BI}^c}{h^2\cos^2\beta} = \frac{1}{2}p_\beta^2\cos^2\beta + \frac{1}{\sqrt{3}}p_\beta + \frac{p_\alpha^2}{2\sin^2\beta\sin^2\gamma} + \frac{p_\gamma^2}{2\sin^2\beta} + U_\phi(\beta,\gamma)e^{-2m} \\
        &U_\phi(\beta,\gamma) = V_\phi(\beta,\gamma)e^{-\frac{2}{\sqrt{3}}|\tan\beta|}\sec^2\beta \\
    \end{aligned}
\end{equation}
The equations of motion generated by the shape space Bianchi I Hamiltonian \ref{shape_H_BI} are 
\begin{equation}
    \label{Bianchi_I_Matter_eom}
    \begin{aligned}
        \dot{\alpha} &= \frac{p_\alpha}{\sin^2\beta\sin^2\gamma} \\
        \dot{\beta} &= p_\beta\cos^2\beta + \frac{1}{\sqrt{3}} \\
        \dot{\gamma} &= \frac{p_\gamma}{\sin^2\beta} \\
        \dot{p}_\alpha &= \left(2sU_\phi p_\alpha - \partial_\alpha U_\phi\right)e^{-2m} \\
        \dot{p}_\beta &= p_\beta^2\cos\beta\sin\beta + \frac{p_\alpha^2\cos\beta}{\sin^3\beta\sin^2\gamma} + \frac{p_\gamma^2\cos\beta}{\sin^3\beta} + \left(2sU_\phi p_\beta - \partial_\beta U_\phi\right)e^{-2m} \\
        \dot{p}_\gamma &= \frac{p_\alpha^2\cos\gamma}{\sin^2\beta\sin^3\gamma} + \left(2sU_\phi p_\gamma - \partial_\gamma U_\phi \right)e^{-2m} \\
        \dot{m} &= -s\left(\frac{1}{\sqrt{3}}p_\beta + 2U_\phi e^{-2m}\right) 
    \end{aligned}
\end{equation}
The right hand sides of the equations of motion \ref{Bianchi_I_Matter_eom} are locally Lipschitz continuous around $\beta = \pi/2$ provided that the field potential does not grow faster than the exponential suppression in $U_\phi(\beta,\gamma)$ i.e.
\begin{equation}
    \lim_{\beta\rightarrow\frac{\pi}{2}}\left(V_\phi(\beta,\gamma)e^{-\frac{2}{\sqrt{3}}|\tan\beta|}\right) = 0
\end{equation}
Since the right hand-sides of the equations of motion are locally Lipschitz continuous, by Picard-Lindelöf there exists a unique local solution to the initial value problem.

In section 2.3.2 the contact Hamiltonian and contact form for a Bianchi cosmology with non-zero shape potential and field potential is shown to be \ref{Contact_Bianchi_Scalar_H}
\begin{equation}
    \begin{aligned}
        \mathcal{H}^c &= -\frac{1}{6}h^2 + \frac{1}{2}\left(p_x^2 + p_y^2\right) + \frac{1}{2}p_\phi^2 + V_s(x,y) + \frac{V(\phi)}{p_k} \\
        \eta &= -dh -\frac{1}{4}p_kdk + p_\phi d\phi + p_xdx + p_ydy
    \end{aligned}
\end{equation}
This had to be handled differently to the vacuum Bianchi case as the shape potential and field potential terms scale differently with $\nu$. We now show the projection of this system onto shape space. In doing so, we will see how the well known conditions for Bianchi IX quiescence translate to the shape space representation.
Beginning with the gnomonic projection of $(\phi,x,y)$ onto $S^3$, and writing the momenta in terms of polar coordinates
\begin{equation}
    \begin{pmatrix}
        \phi \\
        x \\
        y \\
    \end{pmatrix} = |\tan\beta|
    \begin{pmatrix}
        \cos\gamma \\
        \sin\gamma\cos\alpha \\
        \sin\gamma\sin\alpha
    \end{pmatrix}, \quad 
    \begin{pmatrix}
        p_\phi \\
        p_x \\
        p_y
    \end{pmatrix} = p
    \begin{pmatrix}
        \cos\theta \\
        \sin\theta\cos\psi \\
        \sin\theta\sin\psi
    \end{pmatrix}
\end{equation}
along with the transformation
\begin{equation}
\label{BIX_m_h}
    h = se^{m + \frac{1}{\sqrt{3}}|\tan\beta|}, \quad s = \text{Sign}(\tan\beta)
\end{equation}
and finally we define the variables $\chi$ and $\Omega$ as 
\begin{equation}
    \chi = \frac{p}{|h|}, \quad \Omega = -\frac{p_k}{4|h|}
\end{equation}
the rescaled contact form then becomes
\begin{equation}
    \begin{aligned}
        \frac{\eta}{sh} &= -sdm -a\sec^2\beta d\beta + \Omega dk + \chi \cos\theta d\phi + \chi\sin\theta\cos\psi dx + \chi\sin\theta\sin\psi dy \\
        &= -sdm + \Omega dk + p_\alpha d\alpha + p_\beta d\beta + p_\gamma d\gamma
    \end{aligned}
\end{equation}
We identify the shape space momenta as 
\begin{equation}
    \begin{aligned}
        &p_\alpha = \chi|\tan\beta|\sin\gamma\sin\theta\sin(\psi-\alpha) \\
        &p_\beta = \sec^2\beta\left[s\chi\left(\cos\gamma\cos\theta + \sin\gamma\sin\theta\cos(\psi-\alpha)\right)-a\right] \\
        &p_\gamma = \chi|\tan\beta|\left[-\sin\gamma\cos\theta + \cos\gamma\sin\theta\cos(\psi-\alpha)\right]
    \end{aligned}
\end{equation}
The re-scaled shape space Hamiltonian is therefore
\begin{equation}
\label{BIX_matter_H}
    \begin{aligned}
        &\mathcal{H} = \frac{\mathcal{H}^c}{h^2\cos^2\beta} = \frac{1}{2}p_\beta^2\cos^2\beta + \frac{1}{\sqrt{3}}p_\beta + \frac{p_\alpha^2}{2\sin^2\beta\sin^2\gamma} + \frac{p_\gamma^2}{2\sin^2\beta} + U(\alpha,\beta,\gamma)e^{-2m} + \frac{U_\phi(\beta,\gamma)}{s\Omega}e^{-3m} \\
        &U(\alpha,\beta,\gamma) = V_s(\alpha,\beta,\gamma)e^{-\frac{2}{\sqrt{3}}|\tan\beta|}\sec^2\beta \\
        &U_\phi(\beta,\gamma) = V_\phi(\beta,\gamma)e^{-\sqrt{3}|\tan\beta|}\sec^2\beta
    \end{aligned}
\end{equation}
The equations of motion generated by the shape space Hamiltonian are 
\begin{equation}
\label{BIX_matter_eom}
    \begin{aligned}
        \dot{\alpha} &= \frac{p_\alpha}{\sin^2\beta\sin^2\gamma} \\
        \dot{\beta} &= p_\beta\cos^2\beta + \frac{1}{\sqrt{3}} \\
        \dot{\gamma} &= \frac{p_\gamma}{\sin^2\beta} \\
        \dot{k} &= -\frac{U_\phi}{s\Omega^2}e^{-3m} \\
        \dot{p}_\alpha &= \left[\left(\frac{3U_\phi}{\Omega}e^{-m} + 2sU\right)p_\alpha - \partial_\alpha U\right]e^{-2m} \\
        \dot{p}_\beta &= p_\beta^2\cos\beta\sin\beta + \frac{p_\alpha^2\cos\beta}{\sin^3\beta\sin^2\gamma} + \frac{p_\gamma^2\cos\beta}{\sin^3\beta} + \left[\left(\frac{3U_\phi}{\Omega}e^{-m} + 2sU\right)p_\beta - \partial_\beta U - \frac{e^{-m}}{s\Omega}\partial_\beta U_\phi\right]e^{-2m} \\
        \dot{p}_\gamma &= \frac{p_\alpha^2\cos\gamma}{\sin^2\beta\sin^3\gamma} + \left[\left(\frac{3U_\phi}{\Omega}e^{-m} + 2sU\right)p_\gamma - \partial_\gamma U - \frac{e^{-m}}{s\Omega}\partial_\gamma U_\phi\right]e^{-2m} \\
        \dot{\Omega} &= \left(3U_\phi e^{-m} + 2s\Omega U\right)e^{-2m} \\
        \dot{m} &= -s\left(\frac{1}{\sqrt{3}}p_\beta + 2Ue^{-2m}\right) - \frac{3U_\phi}{\Omega}e^{-3m}
    \end{aligned}
\end{equation}
The right-hand sides of the equations of motion \ref{BIX_matter_eom} will be locally Lipschitz continuous around $\beta = \pi/2$ provided that $|\Omega| > 0$, and the potentials $U$ and $U_\phi$ and their derivatives are locally bounded around $\beta = \pi/2$.
One see's that the conditions on the potentials will be satisfied if $V_s$ and $V_\phi$ do not grow faster than their exponential suppression. In particular, the condition
\begin{equation}
    \lim_{\beta \rightarrow \frac{\pi}{2}}\left(V_\phi(\beta,\gamma)e^{-\sqrt{3}|\tan\beta|}\right) = 0
\end{equation}
is the direct translation to shape space of the field potential quiescence condition for mixmaster behaviour to end in a finite number of bounces in Bianchi IX.

In section \ref{VB} it was established that in the vacuum Bianchi IX model, for any given value of $\alpha$, there was always at least one function $f_i(\alpha)$ such that $f_i(\alpha) - 2/\sqrt{3}\geq 0$ and thus the potential 
\begin{equation}
    U(\alpha,\beta) = sec^2\beta\sum_i c_ie^{\left(f_i(\alpha)-\frac{2}{\sqrt{3}}\right)|\tan\beta|}
\end{equation}
diverges at $\beta = \pi/2$.
In the case of Bianchi IX + a scalar field, the anistropy parameters are given by
\begin{equation}
    x = |\tan\beta|\sin\gamma\cos\alpha, \quad y = |\tan\beta|\sin\gamma\sin\alpha
\end{equation}
and the potential term in the shape space Hamiltonian \ref{BIX_matter_H} becomes 
\begin{equation}
\label{U_BIX}
    U(\alpha,\beta,\gamma) = \sec^2\beta\sum_i c_ie^{\left(f_i(\alpha)\sin\gamma-\frac{2}{\sqrt{3}}\right)|\tan\beta|}
\end{equation}

where the functions $f_i(\alpha)$ as defined as they were in eq. \ref{f_i} with coefficients $c_i$. One can now use the additional degree of freedom $\gamma \in [0,\pi]$  to ensure exponential suppression at $\beta = \pi/2$.
Define the set
\begin{equation}
    M = \left\{\max_{0\leq \alpha \leq 2\pi}f_i(\alpha) | i = 1,2...,6\right\}
\end{equation}

Then the exponential suppression will be retained for any $\alpha \in [0,2\pi]$ if 
\begin{equation}
    \sin\gamma < \frac{2}{\sqrt{3}\sup(M)} 
\end{equation}
The form of all functions $f_i(\alpha)$ are known and the supremum of M is $\sup(M) = 4/\sqrt{3}$. Thus we have the quiescence condition
\begin{equation}
    \sin\gamma < \frac{1}{2}
\end{equation}

The conditions under which unique solutions passing through the big bang for the various Bianchi (and FLRW) cosmologies now been established. In the next section we present numerical solutions to the equations of motion for various example models.

\section{Numerical Simulations}
\label{numsect}
In this section we will present numerical solutions to the equations of motion for a variety of the models discussed in this paper. It will be shown that there exist solutions who pass smoothly through the big bang. Under the gnomonic projection, the initial singularity of General Relativity is mapped to $\beta = \pi/2$. The contact Hamiltonians are defined on a contact manifold $T^*Q \times \mathbf{R}$ where $T^*Q$ is an 2n-dimensional cotangent bundle. The gnomonic projection maps the dynamical variables on $T^*Q$ to $S^n$.

\subsection{FLRW + 1 Scalar Field}
The simplest model we can consider is that of FLRW + 1 scalar field. The gnomonic projection in this case maps the dynamical variable $\phi$ onto $S^1$. The shape space Hamiltonian for this model is given by eq.\ref{H_FLRW_1}. 
\begin{equation}
    \begin{aligned}
        &\mathcal{H} = \frac{1}{2}p_\beta^2\cos^2\beta + \frac{\sqrt{3}}{2}p_\beta + U(\beta)e^{-2m} \\
        &U(\beta) = V_\phi(\beta)e^{-\sqrt{3}|\tan\beta|}\sec^2\beta
    \end{aligned}
\end{equation}
In section 3.1 it was shown that the free-field solution is exactly solvable, constant, everywhere-zero momentum $p_\beta$ and linear solution for $\beta(t)$
\begin{equation}
    \beta(t) = \beta_0 + \frac{\sqrt{3}}{2}t
\end{equation}
The system reaches the Big Bang in finite coordinate time 
\begin{equation}
\label{Free_Field_ts}
    t_s = \frac{1}{\sqrt{3}}(\pi-2\beta_0)
\end{equation}
We will consider the numerical solution for a scalar field in a harmonic potential. In the shape space representation this potential is
\begin{equation}
    V(\phi) = \frac{1}{2}\phi^2 = \frac{1}{2}\tan^2\beta
\end{equation}
The potential function $U(\beta)$ is locally Lipschitz continuous around $\beta = \pi/2$, and so Picard-Lindelöf is expected to hold here. 

\clearpage

\begin{figure}[h]
    \centering
    \includegraphics[scale = 0.7]{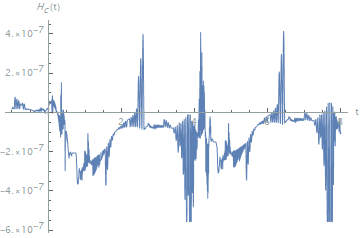}
    \caption{Plot of the numerical solution Hamiltonian for FLRW + Harmonic Potential over the time domain $t\in[-1,8]$. }
    \label{FLRW_1_PhiSquared_H}
\end{figure}
In figure \ref{FLRW_1_PhiSquared_H} we plot the numerical solution Hamiltonian for the FLRW + scalar field in a harmonic potential model. The Hamiltonian constraint requires $\mathcal{H}:=0$, this is satisfied within machine precision, in this case on the order of $10^{-7}$. The initial conditions are chosen at $t = 0$ as

\begin{equation}
\label{PhiSquared_inits}
    \begin{aligned}
        &\beta(0) = \frac{\pi}{8}, \quad p_\beta = \frac{\sqrt{3}}{2}\sec^2\beta\left[-1 + \sqrt{1-\frac{8}{3}U(\beta)e^{-2m}\cos^2\beta}\right] \\
        &m(0) = 0
    \end{aligned}
\end{equation}
where the $p_\beta$ momentum is determined by the Hamiltonian constraint $\mathcal{H}:=0$.
\begin{figure}[h]
    \centering
    \includegraphics[scale = 0.7]{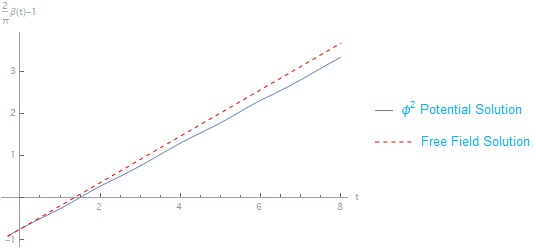}
    \caption{Plot of $\frac{2}{\pi}\beta(t)-1$ for both the free field (red, dashed) and harmonic potential (blue) models. Showing explicitly that both solutions pass through $\beta = \pi/2$ at approximately $t_s = 1.5$. }
    \label{Phi_Squared_B_Check}
\end{figure}

In figure \ref{Phi_Squared_B_Check} we plot the numerical solution of $\beta(t)$, transformed so that it intercepts the t-axis exactly when $\beta = \pi/2$. We see explicitly that the solutions pass through $\beta = \pi/2$ in finite coordinate time. As shown in eq \ref{Free_Field_ts}, the time at which the free field system passes through the big bang can be calculated analytically, as $t_s \approx 1.4$ for these particular initial conditions. The harmonic potential solutions and free-field solutions lie very close together, particularly near $\beta = \pi/2$  which is to be expected as the potential becomes exponentially suppressed close to the big bang and thus the harmonic potential model becomes approximately free. 

The other dynamical variables for this model are the momentum $p_\beta$ and frictional global coordinate $m(t)$. In the free field case the momentum is everywhere zero and $m(t)$ is a constant, which for the initial conditions \ref{PhiSquared_inits} is also zero. 

\begin{figure}[h]
    \centering
    \includegraphics[scale = 0.5]{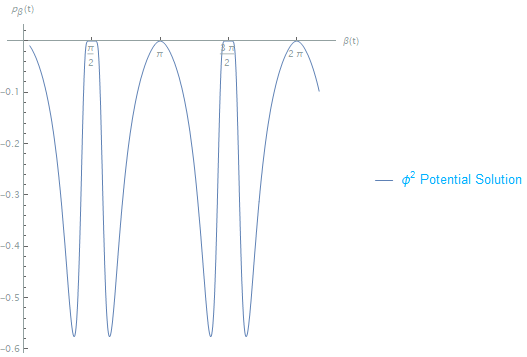}
    \caption{$(\beta,p_\beta)$ phase space plot for the numerical solution of the FLRW + harmonic potential model.}
    \label{FLRW_1_PhiSquared_Phase_B_PB}
\end{figure}
The phase space of the numerical solution to the harmonic potential model is plotted in figure \ref{FLRW_1_PhiSquared_Phase_B_PB}. The numerical solution of $p_\beta(\beta)$ is $\pi$-periodic, with a cyclical structure as the spatial manifold undergoes inversions of orientation and $\text{Sign}(\tan\beta)$ changes between 1 and -1. The momentum also has the expected characteristic exponential suppression where $|\tan\beta| \rightarrow \infty $. Most importantly, the momentum is well defined through $\beta = \pi/2$.

\begin{figure}[h]
    \centering
    \includegraphics[scale = 0.5]{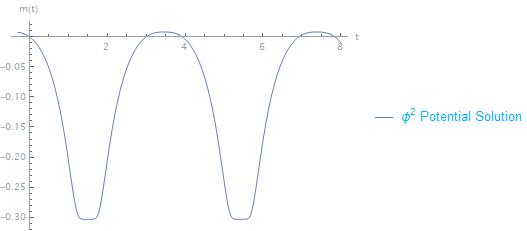}
    \caption{Numerical solution of $m(t)$ for the FLRW + harmonic potential model.}
    \label{FLRW_1_PhiSquared_m}
\end{figure}
Lastly, we plot in figure \ref{FLRW_1_PhiSquared_m} the numerical solution for $m(t)$ for the harmonic potential model. This solution also displays a periodic structure and remains well defined through the big bang.

As well as simple toy models such as a harmonic potential, it is also possible to find numerical solutions for more complicated potentials, provided that they satisfy the conditions for Lipschitz continuity of $U(\beta)$. As an example, consider the Quartic Hilltop  potential 
\begin{equation}
    V_{QH}(\phi) = \Lambda\left[1-\lambda\left(\frac{\phi}{m_{pl}}\right)^4\right]
\end{equation}
which has been of particular interest in inflationary cosmology \cite{Planck:2018jri,Boubekeur_2005,Hoffmann:2021vty,DIMOPOULOS2020135688,Kallosh_2019,Linde:2007fr,Liddle_Lyth_2000}. The parameter $\lambda$ is required to be very small, $\lambda \lesssim 10^{-4}$, in order for the inflationary predictions of the Quartic Hilltop potential to be consistent with 2018 Planck data \cite{DIMOPOULOS2020135688}. $\Lambda$ is a constant energy density scale. In shape space the Quartic Hilltop potential becomes (in units where $m_{pl}=1$)
\begin{equation}
    V_{QH}(\beta) = \Lambda(1-\lambda\tan^4\beta)
\end{equation}
The Quartic Hilltop potential is well known to be unbounded below with no stable vacuum. One stabilised version of the potential which has been investigated in detail is the Quartic Hilltop Squared (QHS) model \cite{Hoffmann:2021vty}, with potential
\begin{equation}
    \begin{aligned}
        &V_{QHS}(\phi) = \Lambda\left[1-\lambda\left(\frac{\phi}{m_{pl}}\right)^4\right]^2 \\
        &V_{QHS}(\beta) - \Lambda\left(1-\lambda\tan^4\beta\right)^2
    \end{aligned}
\end{equation}
For both the Quartic Hilltop and Quartic Hilltop Squared potentials, the potential function $U(\beta)$ is clearly still locally Lipschitz continuous around $\beta = \pi/2$. So there will exist a unique local solution that passes through the big bang. 

\begin{figure}[h]
    \centering
    \includegraphics[scale = 0.7]{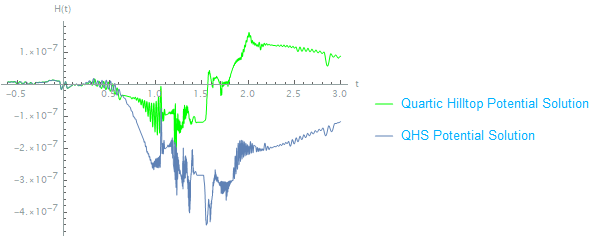}
    \caption{Plot of the numerical solutions Hamiltonian for the FLRW + Quartic Hilltop (green) and Quartic Hilltop Squared (blue) models with initial conditions \ref{PhiSquared_inits}.}
    \label{FLRW_1_QH_H}
\end{figure}
In figure \ref{FLRW_1_QH_H} the numerical solution Hamiltonian for Quartic Hilltop and Quartic Hilltop Squared models, which are approximately zero to within machine precision over the time domain. The initial conditions are taken as those in eq \ref{PhiSquared_inits}, the energy-density scale is chosen as $\Lambda = 1/3$ and the potential parameter is $\lambda = 10^{-2}$. We have chosen a $\lambda$ a few orders of magnitude larger than what is actually required for observational consistency for illustrative purposes that become apparent when one examines the numerical solutions of the dynamical variables.

\clearpage

\begin{figure}[h]
    \includegraphics[scale = 0.7]{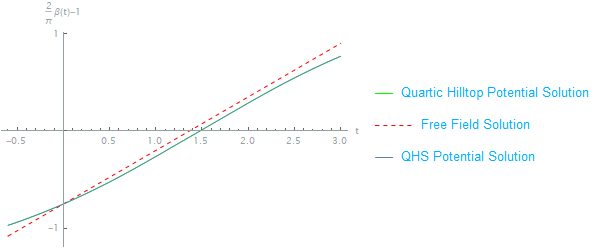}
    \caption{ Plots of $\frac{2}{\pi}\beta(t)-1$ for the free field (red, dashed), Quartic Hilltop (green) and Quartic Hilltop Squared (blue) models, showing explicitly that the system evolves through $\beta = \pi/2$ in finite coordinate time $t_s\approx 1.5$.}
    \label{FLRW_1_QH_B_Check}
\end{figure}
In figure \ref{FLRW_1_QH_B_Check} we plot the numerical solutions for the free-field, Quartic Hilltop and Quartic Hilltop Squared models, again transformed so that they intercept the t-axis exactly when $\beta = \pi/2$ at $t_s \approx 1.5$ . For a small potential parameter $\lambda$ the QH and QHS potentials are approximately equal far from the QHS vacuum expectation value, and thus the solutions lie almost on top of each other. An even smaller value of $\lambda$ closer to what is required for observational consistency with inflationary measurements would force these two solutions closer together.

\begin{figure}[h]
    \centering
    \includegraphics[scale = 0.5]{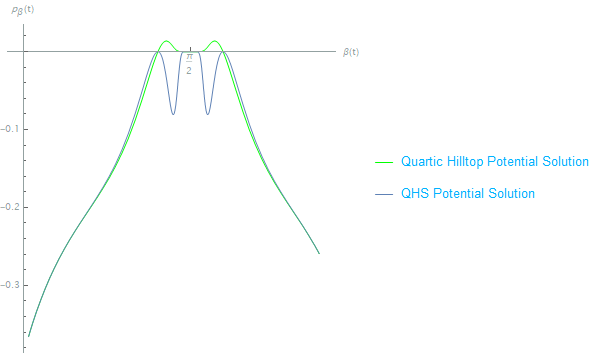}
    \caption{$(\beta,p_\beta)$ phase space plot for the numerical solutions of the Quartic Hilltop (green) and Quartic Hilltop Squared (blue) models.}
    \label{FLRW_1_QH_Phase_B_PB}
\end{figure}

Although the numerical solutions for $\beta(t)$ are almost indistinguishable, the solutions of the momentum $p_\beta(t)$ displayed in the phase space diagram of figure \ref{FLRW_1_QH_Phase_B_PB} are quite distinct, particularly in a neighbourhood around $\beta = \pi/2$. Both momenta go to zero at the big bang and are thus well defined.

\clearpage

\begin{figure}[h]
    \centering
    \includegraphics[scale = 0.5]{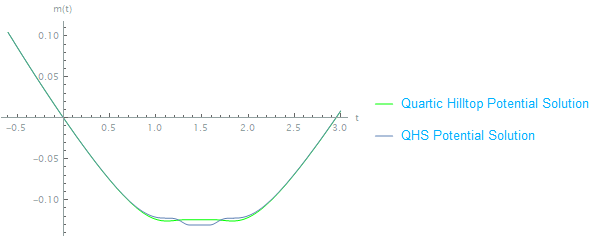}
    \caption{Numerical solution of $m(t)$ for the Quartic Hilltop (green) and Quartic Hilltop Squared (blue) models.}
    \label{FLRW_1_QH_m}
\end{figure}
In figure \ref{FLRW_1_QH_m} we plot the numerical solutions of the final dynamical variable, the frictional global coordinate $m(t)$ for the Quartic Hilltop and Quartic Hilltop Squared models. Just as for the other dynamical variables, the QH and QHS solutions are approximately equal far from the big bang but distinct in a neighbourhood around $t_s \approx 1.5$. For both models $m(t)$ is well defined through the big bang. Thus all dynamical variables, the Hamiltonian and contact form are well defined through the big bang.

\subsection{FLRW + 2 Scalar Fields}
In this section we present examples of a higher dimensional system than that of section 4.1. In the case of FLRW +  2 scalar fields, the shape space projection maps the scalar fields onto $S^2$. The evolution of the fields can be visualised as a path on $S^2$. We showed in section 3.2 that the free-field case was exactly solvable, with the solutions being great circles on $S^2$ \ref{FLRW_2_AB}. The shape space Hamiltonian and equations of motion for FLRW + 2 scalar fields are
\begin{equation}
    \begin{aligned}
        &\mathcal{H} = \frac{\mathcal{H}^c}{h^2\cos^2\beta} = \frac{1}{2}p_\beta^2\cos^2\beta + \frac{\sqrt{3}}{2}p_\beta + \frac{p_\alpha^2}{2\sin^2\beta} + U(\alpha,\beta)e^{-2m} \\
        &U(\alpha,\beta) = V(\alpha,\beta)e^{-\sqrt{3}|\tan\beta|}\sec^2\beta
    \end{aligned}
\end{equation}

\begin{equation}
    \begin{aligned}
        &\dot{\alpha} = \frac{p_\alpha}{\sin^2\beta} \\
        &\dot{\beta} = p_\beta\cos^2\beta + \frac{\sqrt{3}}{2} \\
        &\dot{p}_\alpha = \left(2sUp_\alpha-\partial_\alpha U\right)e^{-2m} \\
        &\dot{p}_\beta = p_\beta^2\cos\beta\sin\beta + \left(2sUp_\beta-\partial_\beta U\right)e^{-2m} \\
        &\dot{m} = -s\left(\frac{\sqrt{3}}{2}p_\beta + 2Ue^{-2m}\right)
    \end{aligned}
\end{equation}

In this example we consider a two-field harmonic potential, which has the following forms in field-space and shape shape respectively
\begin{equation}
    \begin{aligned}
        &V(\phi_1,\phi_2) = \frac{1}{2}\left(\phi_1^2 + \phi_2^2\right) \\
        &V_\phi(\alpha,\beta) = \frac{1}{2}\tan^2\beta
    \end{aligned}
\end{equation}

\clearpage

\begin{figure}[h]
    \centering
    \includegraphics[scale = 0.5]{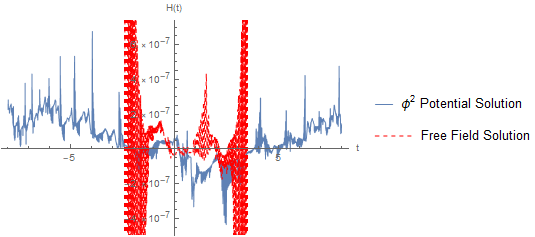}
    \caption{Numerical solution Hamiltonians for the free field (red, dashed) and two-field harmonic potential (blue).}
    \label{FLRW_2_PhiSquared_H}
\end{figure}
In figure \ref{FLRW_2_PhiSquared_H} we plot the numerical solution Hamiltonians for the free-field and harmonic potential models for the initial conditions at $t = 0$
\begin{equation}
    \begin{aligned}
        &\beta(0) = 1, \quad p_\beta(0) = \frac{\sqrt{3}}{2}\sec^2\beta\left[-1 + \sqrt{1-\frac{4}{3}\cos^2\beta\left(\frac{p_\alpha^2}{\sin^2\beta} + 2Ue^{-2m}\right)}\right] \\
        &\alpha(0) = \frac{\pi}{4}, \quad p_\alpha(0) = 0.5 \\
        &m(0) = 0
    \end{aligned}
\end{equation}
Both solutions are stopped just as they become numerically unstable due to the machine precision of the discretized equations of motion becoming insufficient \cite{ardourel:hal-03199593}. This point occurs at different times for each solution, with the harmonic potential solution running for much longer than the free-field solution.

\begin{figure}[h]
    \centering
    \includegraphics[scale = 0.5]{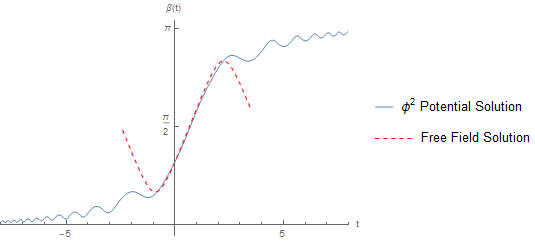}
    \caption{Numerical solutions of $\beta(t)$ for the free field (red, dashed) and two-field harmonic potential (blue).}
    \label{FLRW_2_PhiSquared_B}
\end{figure}

In figure \ref{FLRW_2_PhiSquared_B} we plot the numerical solutions of $\beta(t)$ for the free-field (red, dashed) and harmonic potential (blue) models. Both solutions pass smoothly through $\beta = \pi/2$ in finite coordinate time. We see the characteristic exponential suppression of the potential function $U(\alpha,\beta)$ near $\beta = \pi/2$, making both solutions approximately equal near the big bang.

\clearpage

\begin{figure}[h]
    \centering
    \includegraphics[scale = 0.5]{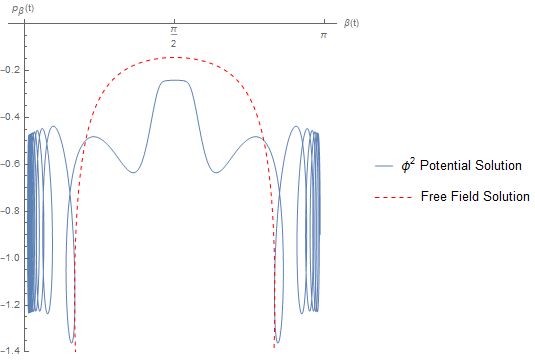}
    \caption{$(\beta,p_\beta)$ phase space plot for the free-field (red, dashed) and harmonic potential (blue) numerical solutions.}
    \label{FLRW_2_Phase_B_PB}
\end{figure}

In figure \ref{FLRW_2_Phase_B_PB} we plot the $(\beta,p_\beta)$ slice of the full 4-dimensional phase space $(\alpha,\beta,p_\alpha,p_\beta)$, for the free field and harmonic potential numerical solutions. In this figure it is easier to see why the Hamiltonians are becoming numerically unstable, particularly for the harmonic potential solutions. As $\beta(t)$ approaches $0,\pi$ the momentum starts to oscillate rapidly. It is unlikely that this is the true behaviour of the system, but rather due to the numerically instability of solving a still system of coupled ODE's. Despite this, the momentum remains well defined at the big bang.

\begin{figure}[h]
\centering
\begin{subfigure}{0.35\textwidth}
    \includegraphics[width=\textwidth]{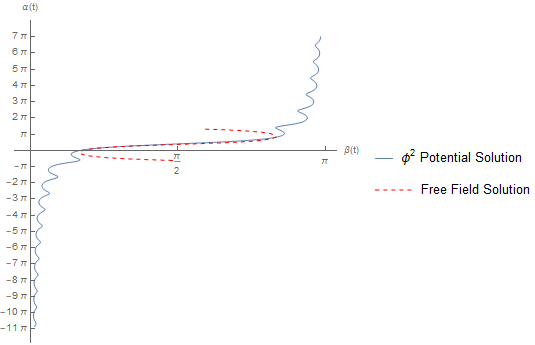}
    \caption{Parametric plot of $(\beta(t),\alpha(t))$ for the free-field (red, dashed) and harmonic potential (blue) numerical solutions.}
    \label{FLRW_2_B_A}
\end{subfigure}
\hfill
\begin{subfigure}{0.35\textwidth}
    \includegraphics[width=\textwidth]{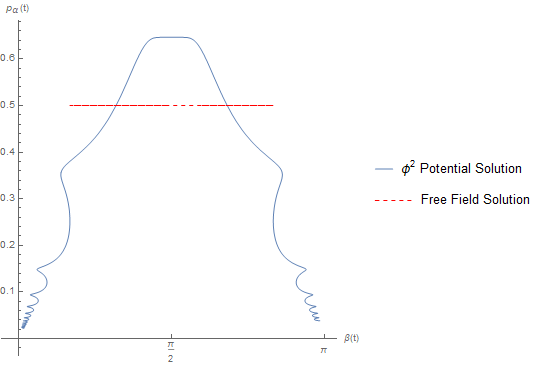}
    \caption{Parametric plot of $(\beta(t),p_\alpha(t))$ for the free-field (red, dashed) and harmonic potential (blue) numerical solutions. }
    \label{FLRW_2_B_PA}
\end{subfigure}
\caption{}
\end{figure}
The parametric plots of $(\beta(t), \alpha(t))$ and $(\beta(t), p_\alpha(t))$ are presented in figures \ref{FLRW_2_B_A} and \ref{FLRW_2_B_PA} respectively for the free-field and harmonic potential numerical solutions. Both dynamical variables $\alpha(t)$ and $p_\alpha(t)$ remain well defined through $\beta = \pi/2$. In particular, in the free-field case $p_\alpha$ is simply a constant. In figure \ref{FLRW_2_B_A} we see that the two solutions are approximately equal near $\beta = \pi/2$, again due to the exponential suppression of the potential close to the big bang. The potential suppression is also manifest in figure \ref{FLRW_2_B_PA}, for the harmonic potential solutions, the $p_\alpha$ becomes approximately constant near $\beta = \pi/2$. 

\clearpage

\begin{figure}[h]
    \centering
    \includegraphics[scale = 0.5]{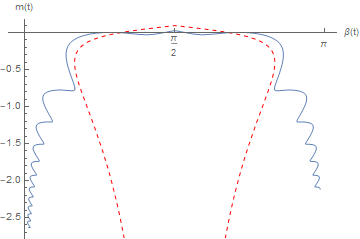}
    \caption{Parametric plot of $(\beta(t), m(t))$ for the free-field (red, dashed) and harmonic potential (blue) numerical solutions.}
    \label{FLRW_2_B_m}
\end{figure}

Lastly in figure \ref{FLRW_2_B_m} we plot the global frictional variable $m(t)$ parameterised by $\beta(t)$ for the free-field and harmonic potential solutions. Both solutions remain well defined through the big bang at $\beta = \pi/2$ and display the characteristic potential suppression. 

\begin{figure}[h]
    \centering
    \includegraphics[scale = 0.5]{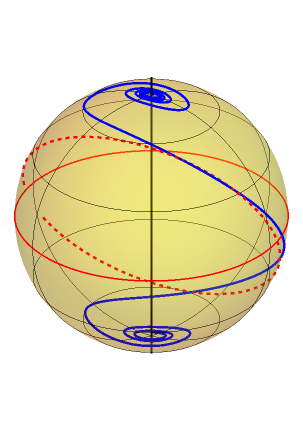}
    \caption{Free-field (red, dashed) and Harmonic potential (blue) numerical solutions of $(\alpha(t),\beta(t))$ plotted as a path on $S^2$. The solid red line is the shape space equator $\beta = \pi/2$, corresponding to the big bang under the gnomonic projection.}
    \label{FLRW_2_S2}
\end{figure}
In figure \ref{FLRW_2_S2} we plot the free-field and harmonic potential solutions $(\alpha,\beta)$ as a path on $S^2$, with the equator at $\beta = \pi/2$ corresponding to the big bang under the shape space gnomonic projection. Recall that we showed explicitly in section 3.2 how the free-field solution is a great circle on shape space \ref{FLRW_2_AB}. This can be seen clearly in figure \ref{FLRW_2_S2}. The great circle is incomplete due to the numerical solution being stopped just before it becomes unstable. Most importantly figure \ref{FLRW_2_S2} shows clearly how the system evolves smoothly through the shape space equator.

\subsection{Bianchi I}
We will now look at numerical results in the shape space representation of the Bianchi I cosmology, which contains flat FLRW geometries at the $S^2$ poles of shape space $\beta = 0, \pi$. The vacuum case simply corresponds to the Kasner solution. The Kasner cosmology is exactly solvable, with the solutions being great circles on $S^2$. In figure \ref{Kasner_S2} below, we present the numerical solution of one such Kasner cosmology, which are parameterised by a single constant momentum $p_\alpha$ \ref{Kasner_GC}.

\begin{figure}[h]
    \centering
    \includegraphics[scale = 0.5]{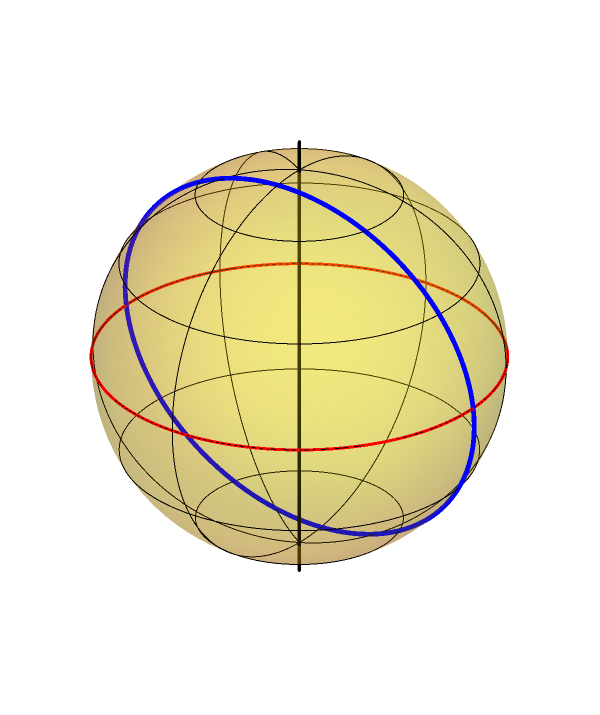}
    \caption{Numerical solution of $(\alpha(t),\beta(t))$ to the vacuum Bianchi I (Kasner) equations of motion \ref{Bianchi_Vacuum_eom}, showing a great circle solution (blue) passing smoothly through the shape space equator (red) $\beta = \pi/2$.}
    \label{Kasner_S2}
\end{figure}

We will now consider in more detail, the case of Bianchi I + a scalar field. In this particular case. The shape space projection now maps the dynamical variables onto $S^3$, as we start with the anistorpy parameters and a single scalar field. We will consider the case of a free-field and harmonic potential. With the Hamiltonian and equations of motion given by eq's \ref{shape_H_BI} and \ref{Bianchi_I_Matter_eom} respectively. Reproduced here for convenience.
\begin{equation}
    \begin{aligned}
        &\mathcal{H}_{BI} = \frac{1}{2}p_\beta^2\cos^2\beta + \frac{1}{\sqrt{3}}p_\beta + \frac{p_\alpha^2}{2\sin^2\beta\sin^2\gamma} + \frac{p_\gamma^2}{2\sin^2\beta} + U_\phi(\beta,\gamma)e^{-2m} \\
        &U_\phi(\beta,\gamma) = V_\phi(\beta,\gamma)e^{-\frac{2}{\sqrt{3}}|\tan\beta|}\sec^2\beta \\
    \end{aligned}
\end{equation}

\begin{equation}
\label{BIeom}
    \begin{aligned}
        \dot{\alpha} &= \frac{p_\alpha}{\sin^2\beta\sin^2\gamma} \\
        \dot{\beta} &= p_\beta\cos^2\beta + \frac{1}{\sqrt{3}} \\
        \dot{\gamma} &= \frac{p_\gamma}{\sin^2\beta} \\
        \dot{p}_\alpha &= \left(2sU_\phi p_\alpha - \partial_\alpha U_\phi\right)e^{-2m} \\
        \dot{p}_\beta &= p_\beta^2\cos\beta\sin\beta + \frac{p_\alpha^2\cos\beta}{\sin^3\beta\sin^2\gamma} + \frac{p_\gamma^2\cos\beta}{\sin^3\beta} + \left(2sU_\phi p_\beta - \partial_\beta U_\phi\right)e^{-2m} \\
        \dot{p}_\gamma &= \frac{p_\alpha^2\cos\gamma}{\sin^2\beta\sin^3\gamma} + \left(2sU_\phi p_\gamma - \partial_\gamma U_\phi \right)e^{-2m} \\
        \dot{m} &= -s\left(\frac{1}{\sqrt{3}}p_\beta + 2U_\phi e^{-2m}\right) 
    \end{aligned}
\end{equation}
The harmonic potential in field space and shape space is given by 
\begin{equation}
    \begin{aligned}
        V(\phi) &= \frac{1}{2}\phi^2 \\
        V_\phi(\beta,\gamma) &= \frac{1}{2}\tan^2\beta\cos^2\gamma
    \end{aligned}
\end{equation}

\clearpage

For this examples, we look for a numerical solution to the equations of motion \ref{BIeom}.  subject to the following initial conditions at $t = 0$. 
\begin{equation}
    \begin{aligned}
        &\alpha(0) = \frac{3}{2}\pi, \quad p_\alpha(0) = 0.1 \\
        &\beta(0) = 1.1, \quad p_\beta(0) = \frac{\sqrt{3}}{2}\sec^2\beta\left[-1 + \sqrt{1-\frac{4}{3}\cos^2\beta\left(\frac{p_\alpha^2}{\sin^2\beta} + 2Ue^{-2m}\right)}\right] \\
        &\gamma(0) = \frac{\pi}{4}, \quad p_\gamma(0) = 1 \\
        &m(0) = 0
    \end{aligned}
\end{equation}
For this section and the following, we provide the Hamiltonian numerical solution plots in appendix \ref{A_H}. In particular, one may find the numerical solution Hamiltonian for Bianchi I + free-field and harmonic potential models in figure \ref{Bianchi_I_H}.

Just as for the FLRW + 2 scalar fields model, we also see in Bianchi I + matter that the free-field solution becomes numerically unstable faster than the harmonic potential model and has to be stopped after a shorter time. We next check the numerical solutions of $\beta(t)$ to confirm that the solutions pass through the big bang in finite coordinate time.

\begin{figure}[h]
    \centering
    \includegraphics[scale = 0.7]{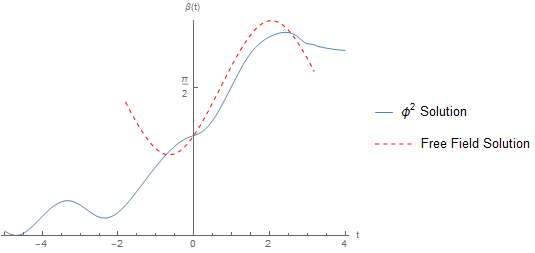}
    \caption{Numerical solution of $\beta(t)$ for the Bianchi I + free-field (red, dashed) and harmonic potential (blue) models.}
    \label{Bianchi_I_B}
\end{figure}
In figure \ref{Bianchi_I_B} we plot the numerical solutions of $\beta(t)$ for the free-field and harmonic potential model. Both solutions pass smoothly through the big bang at $\beta = \pi/2$. In this case the there is a bigger difference between the free-field and harmonic potential solutions over the entire time domain. This is due to the fact that the initial conditions for each solution are slightly different. We choose $p_\beta$ to be set by the Hamiltonian constraint $\mathcal{H}:=0$ (This choice is arbitrary, in principle we could have chosen any of the other dynamical variables). This means that given a set of initial $\alpha, p_\alpha, \gamma, p_\gamma \text{ and } m$, the boundary condition on $p_\beta$ is determined by 
\begin{equation}
    p_\beta = \frac{\sqrt{3}}{2}\sec^2\beta\left[-1 + \sqrt{1-\frac{4}{3}\cos^2\beta\left(\frac{p_\alpha^2}{\sin^2\beta} + 2U(\alpha,\beta)e^{-2m}\right)}\right]
\end{equation}
The potential term $U(\alpha,\beta)$ is non-zero for our chosen initial conditions in the harmonic potential case, and thus the initial $p_\beta$ will be different in the free-field and harmonic potential cases.

\clearpage

\begin{figure}[h]
\centering
\begin{subfigure}{0.45\textwidth}
    \includegraphics[width=\textwidth]{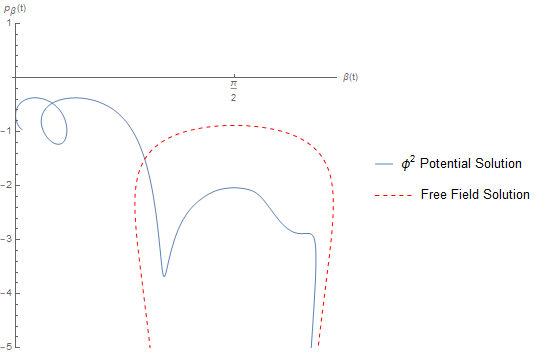}
    \caption{Phase space plot of $(\beta(t),p_\beta(t))$ for the numerical solutions of the Bianchi I + free-field (red, dashed) and harmonic potential (blue) models.}
    \label{Bianchi_I_B_PB}
\end{subfigure}
\hfill
\begin{subfigure}{0.45\textwidth}
    \includegraphics[width=\textwidth]{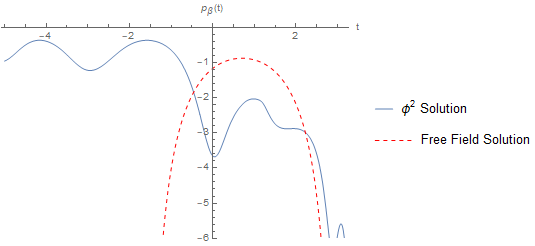}
    \caption{Numerical solution of $p_\beta(t)$ for the numerical solutions of the Bianchi I + free-field (red, dashed) and harmonic potential (blue) models.}
    \label{Bianchi_I_PB}
\end{subfigure}
\caption{}
    \label{}
\end{figure}

In figure \ref{Bianchi_I_B_PB} the numerical solution of $p_\beta$ parameterised by $\beta(t)$ ($(\beta,p_\beta)$ slice of phase space) for the free-field and harmonic potential models. Both solutions, as expected pass smoothly through the big bang at $\beta = \pi/2$. In figure \ref{Bianchi_I_PB} we plot $p_\beta(t)$ parameterised by the time coordinate. This plot shows explicitly that the initial $p_\beta$'s at $t = 0$ are quite different for each model, so it is no surprise that the numerical solutions differ by quite a large amount as compared to some of the previous examples, even near the big bang singularity where the potential is exponentially suppressed.

\begin{figure}[h]
\centering
\begin{subfigure}{0.45\textwidth}
    \includegraphics[width=\textwidth]{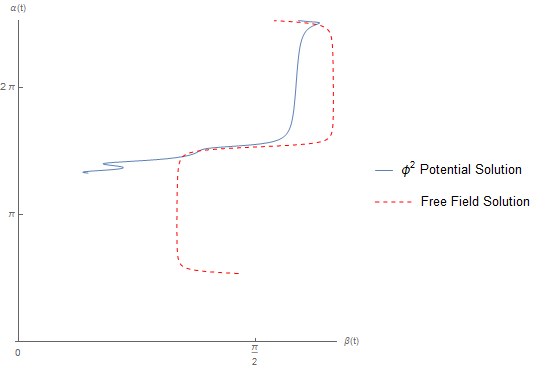}
    \caption{Parametric plot of $(\beta(t),\alpha(t))$ for the Bianchi I + free-field (red, dashed) and harmonic potential (blue) numerical solutions.}
    \label{Bianchi_I_B_A}
\end{subfigure}
\hfill
\begin{subfigure}{0.45\textwidth}
    \includegraphics[width=\textwidth]{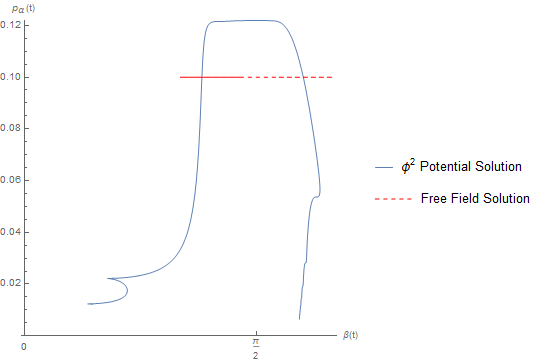}
    \caption{Parametric plot of $(\beta(t),p_\alpha(t))$ for the Bianchi I + free-field (red, dashed) and harmonic potential (blue) numerical solutions.}
    \label{Bianchi_I_B_PA}
\end{subfigure}
\caption{}
    \label{}
\end{figure}
In figures \ref{Bianchi_I_B_A} and \ref{Bianchi_I_B_PA} we plot the numerical solutions of $\alpha$ and $p_\alpha$ respectively, parameterised by $\beta(t)$. As we expect, both solutions pass smoothly though the big bang at $\beta = \pi/2$ along with some other expected features. In figure \ref{Bianchi_I_B_PA}, we see that the free-field $p_\alpha$ momentum is everywhere constant, as it should be since $\dot{p}_\alpha = 0$ everywhere for zero-potential. Likewise, near the big bang where the potential becomes exponentially suppressed, the harmonic potential $p_\alpha$ momentum becomes approximately constant.

\clearpage

\begin{figure}[h]
\centering
\begin{subfigure}{0.45\textwidth}
    \includegraphics[width=\textwidth]{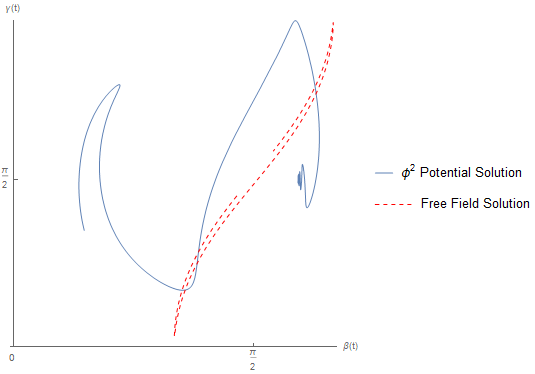}
    \caption{Parametric plot of $(\beta(t),\gamma(t))$ for the Bianchi I + free-field (red, dashed) and harmonic potential (blue) numerical solutions.}
    \label{Bianchi_I_B_G}
\end{subfigure}
\hfill
\begin{subfigure}{0.45\textwidth}
    \includegraphics[width=\textwidth]{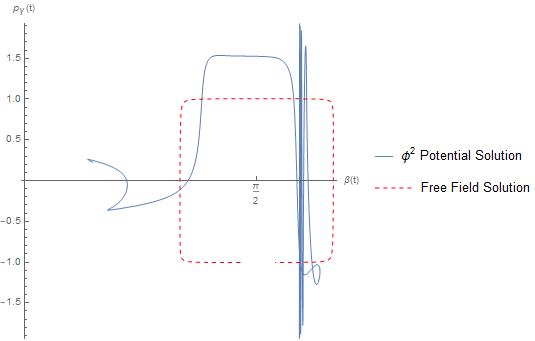}
    \caption{Parametric plot of $(\beta(t),p_\gamma(t))$ for the Bianchi I + free-field (red, dashed) and harmonic potential (blue) numerical solutions.}
    \label{Bianchi_I_B_PG}
\end{subfigure}
\caption{}
    \label{}
\end{figure}

In figures \ref{Bianchi_I_B_G} and \ref{Bianchi_I_B_PG} we plot the numerical solutions of $\gamma$ and $p_\gamma$ respectively, parameterised by $\beta(t)$. We again see that the numerical solutions are well-defined through the big bang at $\beta = \pi/2$.  Additionally, we see a small oscillatory behaviour in $\gamma$ around $\gamma = \pi/2$. The interpretation of this will becomes clear when we look at the original configuration space variables $(\phi,x,y,)$ shortly.

\begin{figure}[h]
    \centering
    \includegraphics[scale = 0.5]{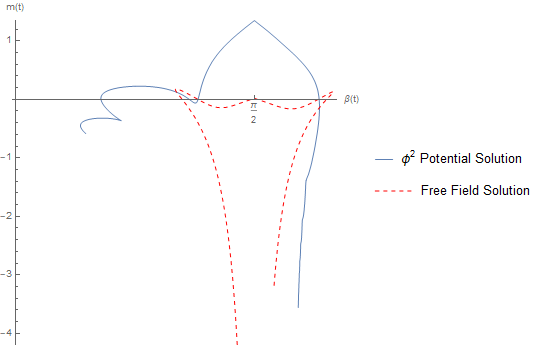}
    \caption{Parametric plot of $(\beta(t),m(t))$ for the Bianchi I + free-field (red, dashed) and harmonic potential (blue) numerical solutions.}
    \label{Bianchi_I_B_m}
\end{figure}
The last dynamical variable to check is the global frictional coordinate $m(t)$, which we plot in figure \ref{Bianchi_I_B_m} for the free-field and harmonic potential models. In both models the numerical solution is well defined through the big bang at $\beta = \pi/2$. Thus all dynamical variables, the Hamiltonian and contact form are well defined through $\beta = \pi/2$ as we expect due to the potential function $U(\alpha,\beta,\gamma)$ being locally Lipschitz continuous around $\beta = \pi/2$. 

In the case of a Bianchi + scalar field cosmology, the shape space projection is a compactification onto $S^3$ (rather than $S^2$ as in the FLRW case), thus we can think of solutions a path on $S^3$. One way to visualise this is to consider constant $\alpha, \beta \text{ or } \gamma$ slices of $S^3$, which are $S^2$ surfaces.

\clearpage

\begin{figure}[h]
    \centering
    \includegraphics[scale = 0.5]{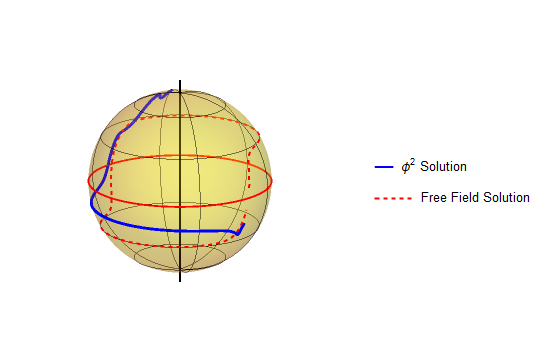}
    \caption{Constant $\gamma$ slice of the Bianchi I + matter $S^3$ shape space. The $(\alpha,\beta)$ solutions for the free-field (red, dashed) and harmonic potential (blue) are plotted as paths on the $S^2$ surface. The solid red line is the $S^2$ slice equator at $\beta = \pi/2$ which corresponds to the GR singularity.}
    \label{Bianchi_I_S2_A_B}
\end{figure}

In figure \ref{Bianchi_I_S2_A_B} we plot a constant $\gamma$ slice of the full $S^3$ shape space with the free-field and harmonic potential paths plotted on the surface. The equator is the line $\beta = \pi/2$ which represents the big bang, both solutions pass smoothly through the shape space equator. However the paths are not symmetric on either side of the equator, despite there being a symmetric (or no) potential, since this is only a constant $\gamma$ slice of the full $S^3$ shape space, on which the path would be symmetric either side of the big bang. Likewise we could also look at constant $\alpha$ and constant $\beta$ slice of shape space.

\begin{figure}[h]
\centering
\begin{subfigure}{0.45\textwidth}
    \includegraphics[width=\textwidth]{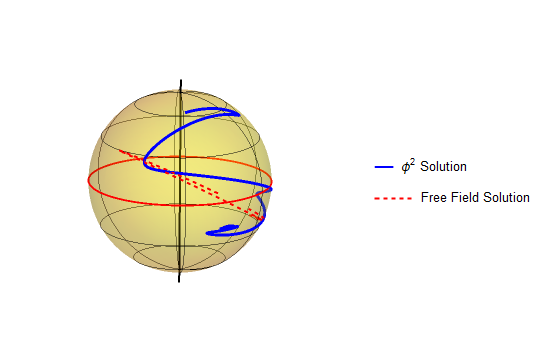}
    \caption{Constant $\alpha$ slice of the Bianchi I + matter $S^3$ shape space. The $(\gamma,\beta)$ solutions for the free-field (red, dashed) and harmonic potential (blue) are plotted as paths on the $S^2$ surface. The solid red line is the $S^2$ slice equator at $\beta = \pi/2$ which corresponds to the GR singularity.}
    \label{Bianchi_I_S2_G_B}
\end{subfigure}
\hfill
\begin{subfigure}{0.45\textwidth}
    \includegraphics[width=\textwidth]{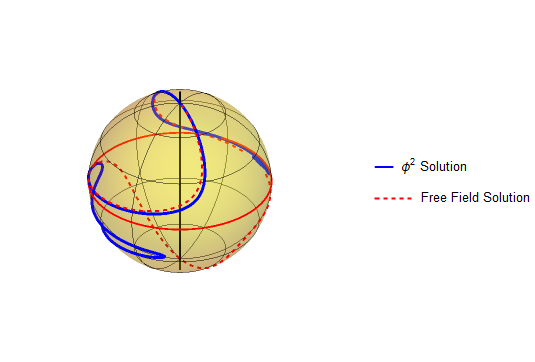}
    \caption{Constant $\beta$ slice of the Bianchi I + matter $S^3$ shape space. The $(\gamma,\alpha)$ solutions for the free-field (red, dashed) and harmonic potential (blue) are plotted as paths on the $S^2$ surface. The solid red line is the $S^2$ slice equator at $\alpha = \pi/2$ which does not correspond to any GR singularity.}
    \label{Bianchi_I_S2_G_A}
\end{subfigure}
\caption{}
    \label{}
\end{figure}

In figures \ref{Bianchi_I_S2_G_B} and \ref{Bianchi_I_S2_G_A}  we plot constant $\alpha$ and $\beta$ slices of the full $S^3$ shape space respectively. In the constant $\alpha$ slice, the solid red line is again the $\beta = \pi/2$ equator, representing the initial GR singularity. The solution path passes smoothly through the big bang equator for both solutions. In the constant $\beta$ slice, the solid red line is the $\alpha = \pi/2$ equator, which does not correspond to any GR singularity and thus the system passing through this line is not of any particular significance in relation to GR singularities. 

In this section we have seen examples of the shape pace dynamical variables and mathematical structures remain well defined through the big bang, as was to be expected from the potential function $U(\alpha,\beta,\gamma)$ satisfying the conditions for Lipschitz continuity. It is still the case however that the original configuration space dynamical variables $\phi,x,y$ are ill-defined though the big bang. One may transform back to the original variables and see this explicitly for the numerical solutions.

\begin{figure}[h]
    \centering
    \includegraphics[scale = 0.5]{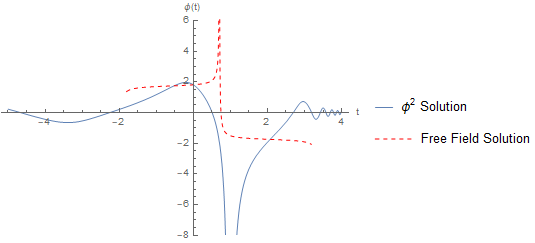}
    \caption{Plot of the original scalar field variable $\phi(t) = |\tan\beta(t)|\cos\gamma(t)$ for the Bianchi I + free-field (red, dashed) and harmonic potential (blue) numerical solutions.}
    \label{Bianchi_I_phi}
\end{figure}

In figure \ref{Bianchi_I_phi} we plot the scalar field under the shape space projection \ref{S3_projection} for the free-field and harmonic potential numerical solutions. Since $\beta(t)$ passes smoothly through $\beta = \pi/2$, the scalar field diverges at the big bang due to the $|\tan\beta|$ factor in the gnomonic projection, but is well defined on either side of the big bang. We also see oscillations of the scalar field around the vacuum $\phi_0 = 0$ on both sides of the GR singularity, although they are more pronounced on the right-half plane. In shape space, these vacuum oscillations manifest as $\gamma$ oscillating around $\gamma = \pi/2$ as seen in figure \ref{Bianchi_I_B_G}.

\begin{figure}[h]
\centering
\begin{subfigure}{0.45\textwidth}
    \includegraphics[width=\textwidth]{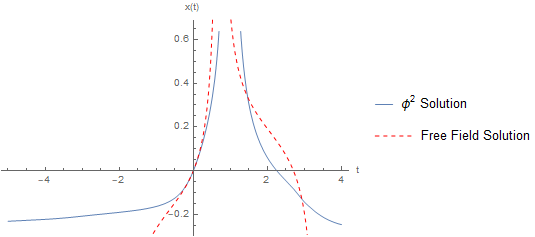}
    \caption{Plot of the original anisotropy parameter $x(t) = |\tan\beta(t)|\sin\gamma(t)\cos\gamma(t)$ for the Bianchi I + free-field (red, dashed) and harmonic potential (blue) numerical solutions.}
    \label{Bianchi_I_x}
\end{subfigure}
\hfill
\begin{subfigure}{0.45\textwidth}
    \includegraphics[width=\textwidth]{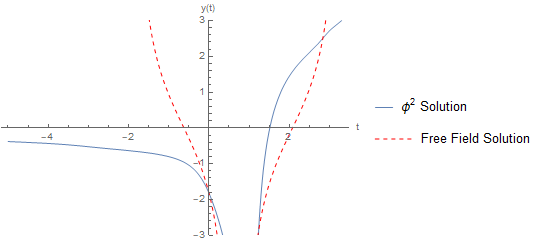}
    \caption{Plot of the original anisotropy parameter $x(t) = |\tan\beta(t)|\sin\gamma(t)\sin\gamma(t)$ for the Bianchi I + free-field (red, dashed) and harmonic potential (blue) numerical solutions.}
    \label{Bianchi_I_y}
\end{subfigure}
\caption{}
    \label{}
\end{figure}

In figures \ref{Bianchi_I_x} and \ref{Bianchi_I_y} we plot the anisotropy parameters $x(t)$ and $y(t)$ given by the shape space projection \ref{S3_projection} for the free-field and harmonic potential numerical solutions. The anisotropy parameters also diverge at the big bang. One would expect from solving the original configuration space equations of motion generated from the Herglotz Lagrangian 
\begin{equation}
    \mathcal{L}^H = \frac{1}{6}h^2 + \frac{1}{2}\left(\dot{x}+\dot{y}\right) + \frac{1}{2}\dot{\phi} - V(\phi)
\end{equation}
that the anisotropy parameters are related to each other by a linear relationship $y(x) = y_0 + C_y x$.
\begin{figure}[h]
    \centering
    \includegraphics[scale = 0.5]{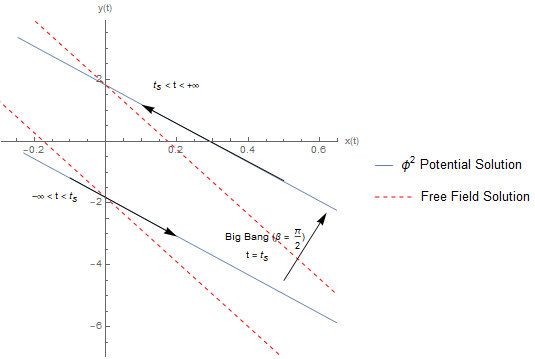}
    \caption{Parameteric plot of $(x(t),y(t))$ for the Bianchi I + free-field (red, dashed) and harmonic potential (blue) numerical solutions. As the system evolves through the Big Bang at $\beta(t_s) = \pi/2$, the spatial manifolds orientation inverts and system moves between straight line branches.}
    \label{Bianchi_I_xy}
\end{figure}
in figure \ref{Bianchi_I_xy} we display the parameteric plot $(x(t),y(t))$ and see that indeed this linear relationship is satisfied. Each solution has two straight line branches, as the system evolves through the big bang the spatial manifold undergoes an inversion of orientation and moves from one branch to the other.

\subsection{Quiescent Bianchi IX}

In this section we look at a numerical example of quiescent Bianchi IX cosmology and show that it continues smoothly through the big bang in the shape space representation. Unlike Bianchi I, the Bianchi IX cosmology has a non-trivial Lie algebra, which leads to generally non-zero shape potential and contains flat FLRW cosmologies. In shape space potential function $U(\alpha,\beta,\gamma)$ for Bianchi IX is given in equation \ref{U_BIX}, with the functions $f_i(\alpha)$ defined in equation \ref{f_i}. The flat FLRW-cosmologies are contained at $\beta = 0,\pi$ and $\gamma = 0,\pi$. However the time parameterisation we have chosen, implicit in the re-scaling of the contact Hamiltonian by $h^{-2}\sec^2\beta$ is not suited for investigating these sub-geometries as the Hamiltonian is undefined for $\beta,\gamma = 0,\pi$.

We will consider Bianchi IX cosmologies minimally coupled to a scalar field in the free-field (massless) and harmonic potential cases. As we have already shown, in the case of a everywhere-zero field potential, the shape space Hamiltonian is
\begin{equation}
\label{BIX_H_eq}
    \begin{aligned}
        &\mathcal{H} = \frac{1}{2}p_\beta^2\cos^2\beta + \frac{1}{\sqrt{3}}p_\beta + \frac{p_\alpha^2}{2\sin^2\beta\sin^2\gamma} + \frac{p_\gamma^2}{2\sin^2\beta} + U(\alpha,\beta,\gamma)e^{-2m} \\
        &U(\alpha,\beta,\gamma) = \sec^2\beta\sum_i c_ie^{\left(f_i(\alpha)\sin\gamma-\frac{2}{\sqrt{3}}\right)|\tan\beta|} \\
    \end{aligned}
\end{equation}
In section 3.5 we showed that, including a generally non-zero field potential changes the shape space Hamiltonian to 
\begin{equation}
    \begin{aligned}
        &\mathcal{H} = \frac{1}{2}p_\beta^2\cos^2\beta + \frac{1}{\sqrt{3}}p_\beta + \frac{p_\alpha^2}{2\sin^2\beta\sin^2\gamma} + \frac{p_\gamma^2}{2\sin^2\beta} + U(\alpha,\beta,\gamma)e^{-2m} + \frac{U_\phi(\beta,\gamma)}{s\Omega}e^{-3m} \\
        &U(\alpha,\beta,\gamma) = \sec^2\beta\sum_i c_ie^{\left(f_i(\alpha)\sin\gamma-\frac{2}{\sqrt{3}}\right)|\tan\beta|}\\
        &U_\phi(\beta,\gamma) = V_\phi(\beta,\gamma)e^{-\sqrt{3}|\tan\beta|}\sec^2\beta
    \end{aligned}
\end{equation}
as additional degrees of freedom $(k,\Omega)$ are required in order to retain the original scaling symmetry of the action. We consider the numerical solutions to the equations of motion \ref{BIX_matter_eom} subject to the following initial conditions at $t = 0$. Simply for demonstrative purposes, we have chosen to use the Hamiltonian constraint to determine the initial condition on $p_\alpha$, rather than $p_\beta$ as in previous sections.
\begin{equation}
    \begin{aligned}
        &\alpha(0) = \frac{1}{2}\pi, \quad p_\alpha(0) = |\sin\beta\sin\gamma|\sqrt{-2\left(\frac{1}{2}p_\beta^2\cos^2\beta + \frac{1}{2}\frac{p_\gamma^2}{2\sin^2\beta} + \frac{1}{\sqrt{3}}p_\beta + Ue^{-2m} + \frac{U_{\phi}}{s\Omega}e^{-3m}\right)} \\
        &\beta(0) = \frac{1}{8}\pi, \quad p_\beta(0) -0.5 \\
        &\gamma(0) = \frac{1}{8}\pi, \quad p_\gamma(0) = 0.1 \\
        &m(0) = 0 \\
        &k(0) = 0, \quad \Omega(0) = -1
    \end{aligned}
\end{equation}

\begin{figure}[h]
    \centering
    \includegraphics[scale = 0.5]{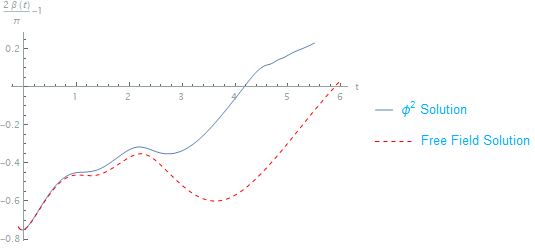}
    \caption{Plots of $\frac{2}{\pi}\beta(t)-1$ for the Bianchi IX $\beta(t)$ numerical solutions, including the free-field (red, dashed) and harmonic potential (blue) models. }
    \label{BIX_B}
\end{figure}
In figure \ref{BIX_B} we plot $\frac{2}{\pi}\beta(t)-1$ where $\beta(t)$ is the Bianchi IX numerical solution for the free-field and Harmonic potential models, showing explicitly that the system passes through the Big Bang at $\beta = \pi/2$. In the free-field case, the discretised system equations of motion become numerically unstable shortly after passing through the Big Bang.

\begin{figure}[h]
    \centering
    \includegraphics[scale = 0.5]{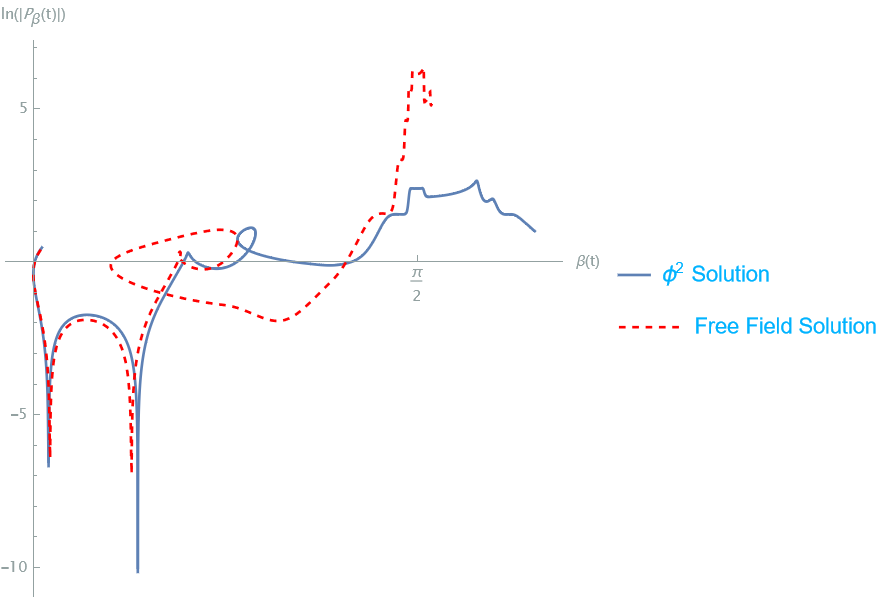}
    \caption{Plots of $\ln|p_\beta(t)|$ parameterised by $\beta(t)$ for the Bianchi IX $p_\beta(t)$ numerical solutions, including the free-field (red, dashed) and harmonic potential (blue) models.}
    \label{BIX_PB}
\end{figure}
In figure \ref{BIX_PB} we plot $\ln|p_\beta(t)|$ parameterised by $\beta(t)$, one can see clearly that in both cases the $p_\beta$ momentum is well defined through $\beta = \pi/2$. In this particular case we choose to plot the logarthim as the free-field and harmonic potential solutions of $p_\beta$ differ by approximately two orders of magnitude near the big bang.

\begin{figure}[h]
\centering
\begin{subfigure}{0.45\textwidth}
    \includegraphics[width=\textwidth]{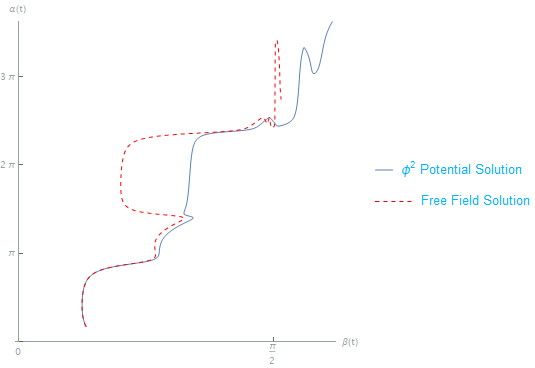}
    \caption{Parametric plot of $(\beta(t),\alpha(t))$ for the Bianchi IX + free-field (red, dashed) and harmonic potential (blue) numerical solutions.}
    \label{BIX_A}
\end{subfigure}
\hfill
\begin{subfigure}{0.45\textwidth}
    \includegraphics[width=\textwidth]{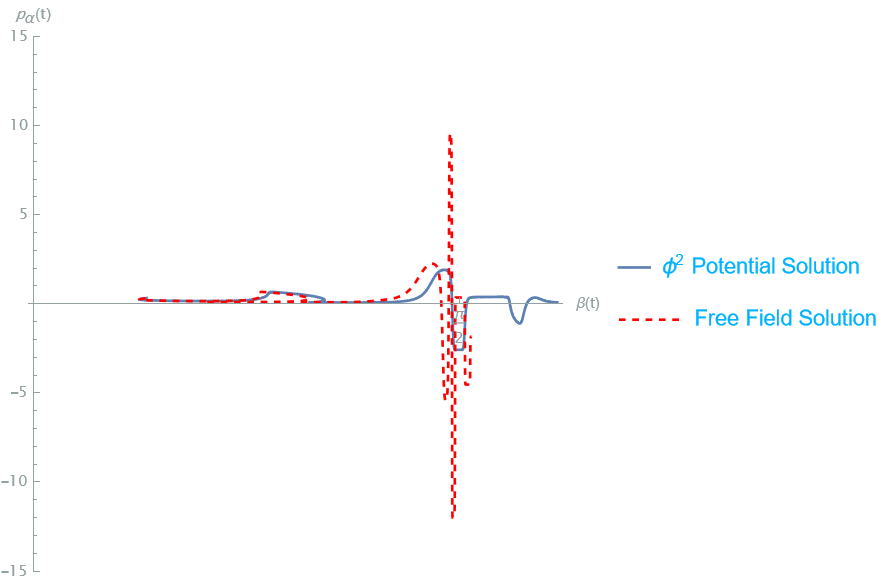}
    \caption{Parametric plot of $(\beta(t),p_\alpha(t))$ for the Bianchi IX + free-field (red, dashed) and harmonic potential (blue) numerical solutions.}
    \label{BIX_PA}
\end{subfigure}
\caption{}
    \label{}
\end{figure}
In figures \ref{BIX_A} and \ref{BIX_PA} we plot the numerical solutions of the shape space coordinate $\alpha$ and associated momentum $p_\alpha$ respectively, parameterised by $\beta$. In both figures we see that the dynamical variables remain well defined through the big bang at $\beta = \pi/2$ as we expect. In the case of Bianchi I, $p_\alpha$ is constant for free-field solutions as the equation of motion for $\dot{p}_\alpha$ contains only terms with factors of $U_\phi$ and $\partial U_\phi$. However in the Bianchi IX case, which contains a shape potential which is not everywhere zero, the equation of motion for $\dot{p}_\alpha$ contains terms which have factors of the shape potential function $U$ as well as the field potential $U_\phi$, so even in the free-field case where $U_\phi = 0$, the momentum $p_\alpha$ is not constant for all time as we see in the numerical solution of figure \ref{BIX_PA}.

\begin{figure}[h]
\centering
\begin{subfigure}{0.45\textwidth}
    \includegraphics[width=\textwidth]{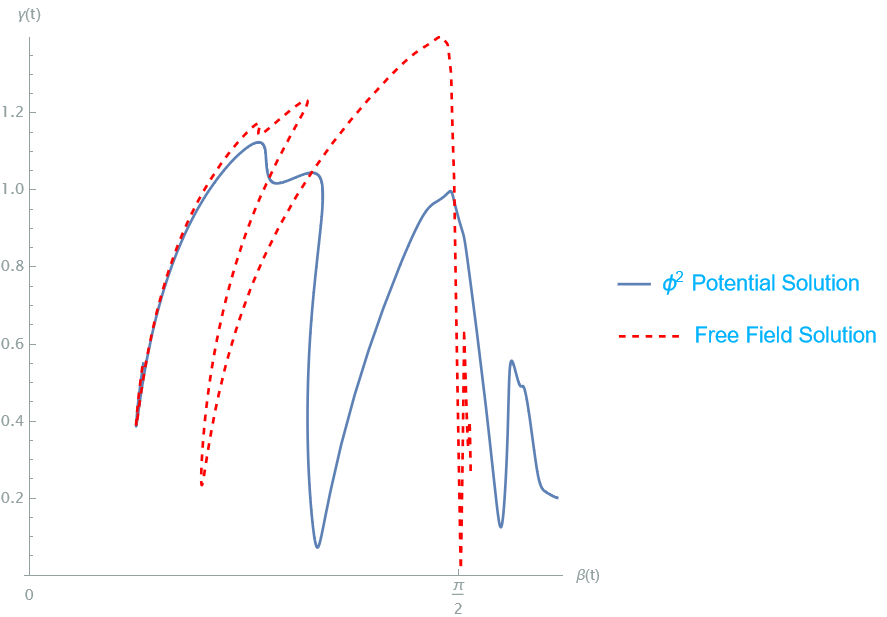}
    \caption{Parametric plot of $(\beta(t),\gamma(t))$ for the Bianchi IX + free-field (red, dashed) and harmonic potential (blue) numerical solutions.}
    \label{BIX_G}
\end{subfigure}
\hfill
\begin{subfigure}{0.45\textwidth}
    \includegraphics[width=\textwidth]{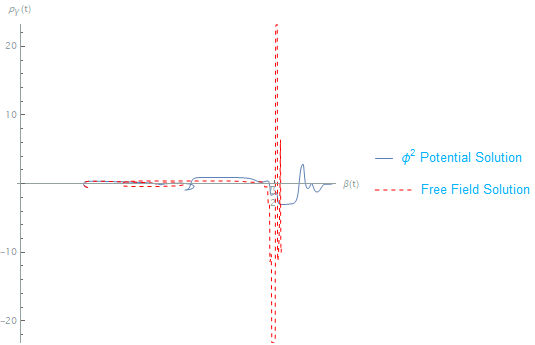}
    \caption{Parametric plot of $(\beta(t),p_\gamma(t))$ for the Bianchi IX + free-field (red, dashed) and harmonic potential (blue) numerical solutions.}
    \label{BIX_PG}
\end{subfigure}
\caption{}
    \label{}
\end{figure}

In figures \ref{BIX_G} and \ref{BIX_PG} we plot the numerical solutions of $\gamma$ and $p_\gamma$ parameterised by $\beta$ for the Bianchi IX + free-field and harmonic potential models. Both dynamical variables remain well defined through the big bang at $\beta = \pi/2$. In addition to the shape degrees of freedom, for the Bianchi IX + matter cosmology we also have the additional degrees of freedom $k$ and its associated momentum $\Omega(t)$ which were introduced to retain the scaling symmetry of the symplectic Lagrangian when there is a field potential present.

\clearpage

\begin{figure}[h]
\centering
\begin{subfigure}{0.45\textwidth}
    \includegraphics[width=\textwidth]{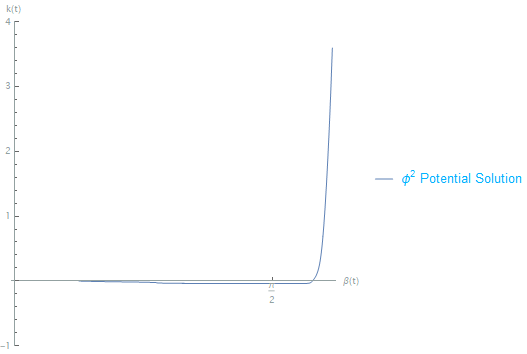}
    \caption{Parametric plot of $(\beta(t),k(t))$ for the Bianchi IX + harmonic potential numerical solution.}
    \label{BIX_k}
\end{subfigure}
\hfill
\begin{subfigure}{0.45\textwidth}
    \includegraphics[width=\textwidth]{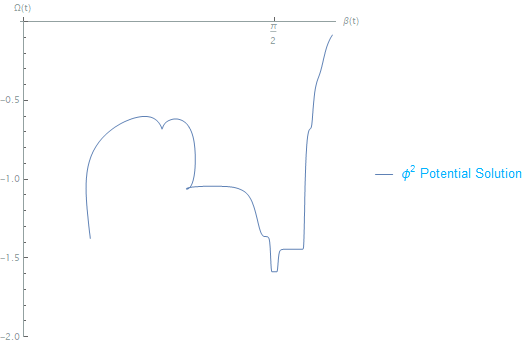}
    \caption{Parametric plot of $(\beta(t),\Omega(t))$ for the Bianchi IX + harmonic potential numerical solutions.}
    \label{BIX_Omega}
\end{subfigure}
\caption{}
    \label{}
\end{figure}
In figures \ref{BIX_k} and \ref{BIX_Omega} we plot the numerical solutions of $k$ and $\Omega$ parameterised by $\beta(t)$ for the Bianchi IX +  harmonic potential model. Both dynamical variables are well defined through the big bang at $\beta = \pi/2$. One one can see from the plot of $(k,\beta)$ that the numerical solution becomes unstable shortly after passing through the big bang as $k\rightarrow\infty$.

\begin{figure}[h]
    \centering
    \includegraphics[scale = 0.5]{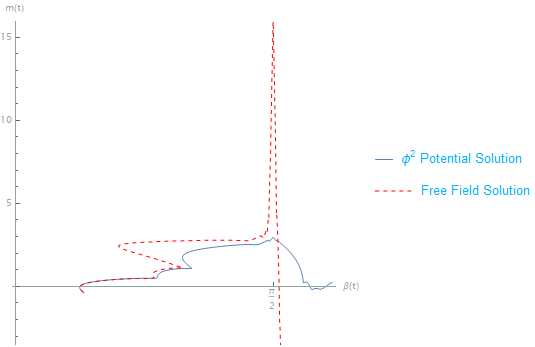}
    \caption{Parametric plot of $(\beta(t),m(t))$ for the Bianchi IX + harmonic potential numerical solution.}
    \label{BIX_m}
\end{figure}

The last dynamical variable is the frictional global coordinate $m(t)$, plotted in figure \ref{BIX_m}
$m$ is also well defined through the big bang and displays the characteristic ``cusp" its time-derivative changes sign as the system passes through the big bang, due to the overall factor of $\text{Sign}(\tan\beta)$ in the equation of motion \ref{BIX_matter_eom}.

Thus we demonstrate that as expected, all dynamical variables, the Hamiltonian and contact form remain well defined as the system passes through the Bianchi IX big bang when the quiescence conditions are satisfied.

\clearpage

Once more can visualise the evolution of the cosmology in terms of a paths on $S^2$ slices of the full $S^3$ shape space.

 \begin{figure}[h]
    \centering
    \includegraphics[scale = 0.5]{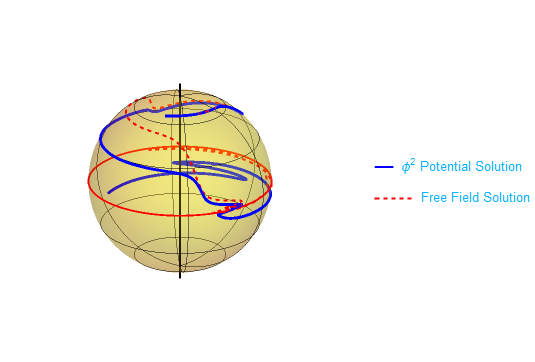}
    \caption{Constant $\gamma$ slice of the Bianchi IX + matter $S^3$ shape space. The $(\alpha,\beta)$ solutions for the free-field (red, dashed) and harmonic potential (blue) are plotted as paths on the $S^2$ surface. The solid red line is the $S^2$ slice equator at $\beta = \pi/2$ which corresponds to the GR singularity.}
    \label{BIX_A_B}
\end{figure}

Figure \ref{BIX_A_B} displays a constant gamma $S^2$ slice of shape space. The free-field and harmonic potential solutions are plotted in red/dashed and blue respectively. Both solution paths pass smoothly through the big bang, represented by the solid red equator at $\beta = \pi/2$.

\begin{figure}[h]
\centering
\begin{subfigure}{0.45\textwidth}
    \includegraphics[width=\textwidth]{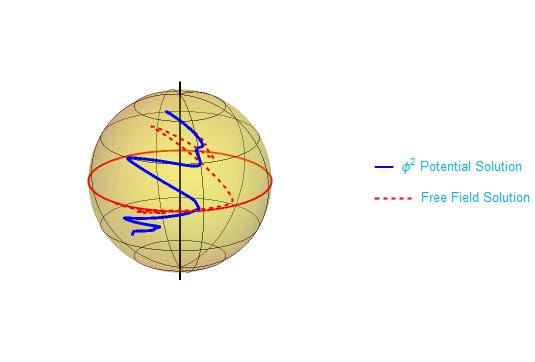}
    \caption{Constant $\alpha$ slice of the Bianchi IX + matter $S^3$ shape space. The $(\gamma,\beta)$ solutions for the free-field (red, dashed) and harmonic potential (blue) are plotted as paths on the $S^2$ surface. The solid red line is the $S^2$ slice equator at $\beta = \pi/2$ which corresponds to the GR singularity.}
    \label{BIX_G_B}
\end{subfigure}
\hfill
\begin{subfigure}{0.45\textwidth}
    \includegraphics[width=\textwidth]{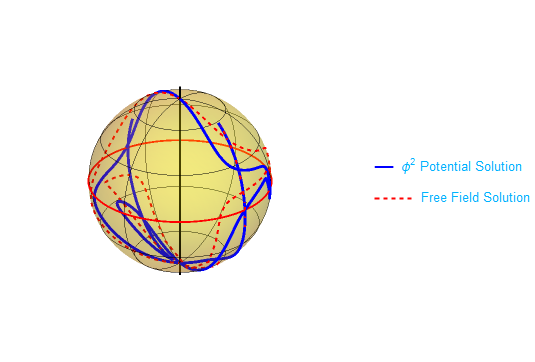}
    \caption{Constant $\beta$ slice of the Bianchi I + matter $S^3$ shape space. The $(\gamma,\alpha)$ solutions for the free-field (red, dashed) and harmonic potential (blue) are plotted as paths on the $S^2$ surface. The solid red line is the $S^2$ slice equator at $\alpha = \pi/2$ which does not correspond to any GR singularity.}
    \label{BIX_A_G}
\end{subfigure}
\caption{}
    \label{}
\end{figure}

Figures \ref{BIX_G_B} and \ref{BIX_A_G} display constant $\alpha$ and constant $\beta$ $S^2$ slices respectively. In figure \ref{BIX_G_B} the equator is still at $\beta = \pi/2$ corresponding to the big bang, which both paths cross smoothly. However in figure \ref{BIX_G_B}, where a constant $\beta$ $S^2$ slice is displayed, the equator is at $\alpha = \pi/2$. which does not correspond to any GR singularity.

Having numerically solved the system in the shape space representation, one can then transform back to the original configuration space dynamical variables of the scalar field $\phi$ and anisotropy parameters $(x,y)$. In particular, it is much easier in this representation to see explicitly, the quiescence behaviour of the cosmology.

\begin{figure}[h]
    \centering
    \includegraphics[scale = 0.5]{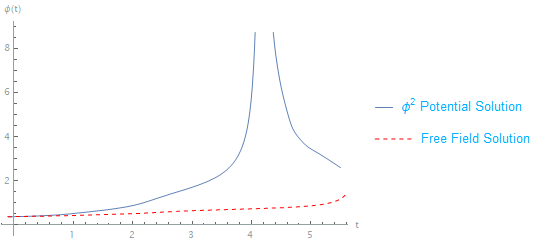}
    \caption{Plot of the original scalar field variable $\phi(t) = |\tan\beta(t)|\cos\gamma(t)$ for the Bianchi IX + free-field (red, dashed) and harmonic potential (blue) numerical solutions.}
    \label{BIX_Phi}
\end{figure}
Firstly we plot in figure \ref{BIX_Phi} the scalar field $\phi = |\tan\beta|\cos\gamma$ for the free-field and harmonic potential numerical solutions, which as expected is undefined at the big bang. Note that the free-field and harmonic potential solutions pass through the big bang at different coordinate times. From figure \ref{BIX_B} one can see that the free-field model passes through the big bang at approximately $t_{\text{s,free}} \approx 5.9$ and the numerical solutions becomes unstable very shortly after, so the divergence is hard to see in figure \ref{BIX_Phi}'s plot of $\phi(t)$. The harmonic potential model passes through the big bang at $t_{\text{s,}\phi^2}\approx 4.2$ and the numerical solution does not becomes unstable until a few coordinate time units later. Thus the divergence in $\phi(t)$ can be seen much clearer for the harmonic potential model in figure \ref{BIX_Phi}.

\begin{figure}[h]
\centering
\begin{subfigure}{0.45\textwidth}
    \includegraphics[width=\textwidth]{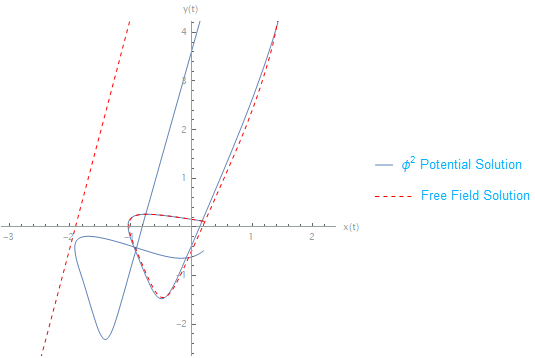}
    \caption{Numerical solution of the anisotropy parameters $(x(t),y(t))$ for the Bianchi IX + free-field (red, dashed) and harmonic potential (blue) models. With the anisotropy parameters given by $x = |\tan\beta|\sin\gamma\cos\alpha$, $y = |\tan\beta|\sin\gamma\sin\alpha$. }
    \label{BIX_x_y}
\end{subfigure}
\hfill
\begin{subfigure}{0.45\textwidth}
    \includegraphics[width=\textwidth]{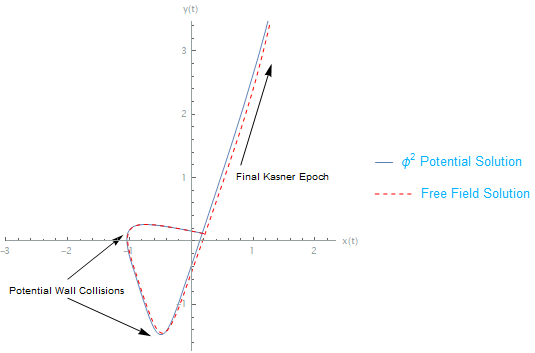}
    \caption{Numerical solution of the anisotropy parameters $(x(t),y(t))$ for the Bianchi IX + free-field (red, dashed) and harmonic potential (blue), over subset of the time domain $t\in[0,t_s]$ ($t_s$ being the time at which each model passes through the big bang singularity). With the anisotropy parameters given by $x = |\tan\beta|\sin\gamma\cos\alpha$, $y = |\tan\beta|\sin\gamma\sin\alpha$.}
    \label{BIX_x_y_partial}
\end{subfigure}
\caption{}
    \label{}
\end{figure}

In figure \ref{BIX_x_y} we plot the anisotropy parameters for the Bianchi IX +  free-field and harmonic potential numerical solutions, over the full time domain. Transforming back to the original configuration space variables makes it much easier to see the quiescent behaviour of the Bianchi IX cosmology. In particular, in figure \ref{BIX_x_y_partial} we plot the anisotropy parameter numerical solutions on a subset of the time domain, from $t = 0$ until $t = t_s$, where $t_s$ is the time at which the models pass through the big bang singularity in shape space. Here we see explicitly Tuab transitions and the bounces of $(x,y)$ off the shape potential wall before being set off one one final Kasner epoch, which the cosmology remains on until it reaches the big bang and spatial contact manifold undergoes and inversion of orientation.

\section{Discussion}
\label{dissect}

The notion of a singularity in General Relativity is a subtle one. Generally speaking, in classical field theories we tend to think of singularities as points at which physical quantities become indeterminate and grow unboundedly. In GR, we solve the Einstein Field Equations for the spacetime manifold and its endowed metric structure. In this sense we can't think of a singularity in general relativity as a point on the manifold where the curvature invariant grows unboundedly, as the Einstein Field Equations need to be solved in an neighbourhood of that point in order to define the spacetime manifold and metric. Therefore the contemporary interpretation of a singularity in GR is that of a boundary to the spacetime manifold on which the curvature invariants become indeterminate \cite{Steinbauer:2022hvq,Rendall:2005nf}. In this work we show how the physical quantities of the cosmological dynamical system and the mathematical structures which generate dynamics (Hamiltonian and contact form) and be smoothly continued across this boundary that stitches two identical, oppositely spatially oriented cosmologies together.

Hawking and Penrose showed that singularities in General Relativity are of a generic nature \cite{1970RSPSA.314..529H,Rendall:2005nf,Steinbauer:2022hvq}, the theory breaks itself with relatively lenient requirements for timelike and null geodesic incompleteness. The fact that singularities occur so generically in GR is the main motivating reason to believe that the theory is incomplete. Many hope that the key to resolving singularities in GR lies at the quantum level, and there is a large school of thought that singularities are where one should look to for effects of quantum gravity. Whilst GR is perturbatively non-renormalisable, non-perturbative approaches like loop quantum gravity \cite{Thiemann:2001gmi} have been successful in resolving spacetime singularities at the quantum level \cite{Li:2023dwy,Gambini:2013hna,Gambini:2022hxr,Ashtekar:2006rx,Bojowald:2001xe,Bojowald:2005epg}. Within the Loop Quantum Gravity framework, the singularities of FLRW (flat, open and closed) \cite{Ashtekar:2006rx,Ashtekar:2006uz,PhysRevD.75.024035,Szulc_2007}, In much of the  literature, it is assumed that quantum effects are necessary to resolve singularities in GR, particularly the initial cosmological singularity due to the small length scales involved \cite{Bojowald:2005epg,Ashtekar_2015}. In the relational dynamics approach however, the Hamiltonian makes no reference to the overall length scale of the universe, which is not a physically observable. The shape dynamical system is subtly different from general relativity in that the fundamental gauge symmetry is that of spatial conformal invariance of the metric, rather than local diffeomorphism invariance \cite{Mercati:2014ama,Barbour:2011dn,Gomes_2013} and its configuration space is an equivalence class of conformal 3-geometries, rather than the space of Riemann 3-geometries.

In this paper we have shown that the singularity can, in a sense, be resolved classically. We have shown how to form a relational, shape space description of flat FLRW, Bianchi I and Quiescent Bianchi IX cosmologies by identifying a scaling symmetry of the Lagrangian associated with the scale factor $\nu$. While the curvature invariant remains divergent at the initial singulairty, in the relational shape space description, determinism of the dynamical system is retained across the big bang. In the literature, homogeneous and black-hole interior spacetime have been continued through their GR singularities \cite{Koslowski:2016hds,Mercati:2021zmv} by forming dimensionless dynamical variables with equations of motion generated by the usual symplectic ADM Hamiltonian. Deterministic continuation through the singularity is established by investigating integrability of the equations of motion. In this work we first form a fully relational contact system by quotienting out the scaling symmetry and forming a symmetry-reduced contact manifold with physically equivalent dynamics to the original symplectic manifold. The equations of motion in the relational system are generated directly from a contact Hamiltonian, which makes no reference to the overall scale of the universe. We also investigate the integrability of this system and have shown that there exist unique solutions to the contact Hamiltonian equations of motion in the shape space description which evolve smoothly through the big bang. Moreover, \emph{the mathematical structures that generate time evolution, namely the Hamiltonian and contact form remain well defined}. This is a key development both in terms of defining the solution space, and when considering potential quantizations of the theory. In this manner we are able to define a classical Hamiltonian that is well defined across the big bang, which may be of particular interest in canonical quantisations of GR, where the Hamiltonian and symplectic structure are of fundamental importance. Shape Dynamics presents a possible new path to a towards a quantum theory of gravity, in particular it offers a solution to the fundamental problem of time in the canonical approach to quantum gravity \cite{anderson2012problem}, in which the Hamiltonian constraint when quantised demands a stationary wavefunction. Shape Dynamics manages to decouple dynamical evolution from scale and admits unambiguous evolution in terms of the York time \cite{Barbour_2014,barbour2013gravitational}.

Alongside showing analytically that cosmological dynamical system on a contact manifold may be continued uniquely through the initial singularity, in this work we have provided complete numerical solutions of the relational equations of motion for each of the cosmologies described above. These numerical solutions show the asymptotic potential-free behaviour near the big bang corresponding to great circles which continue smoothly through the shape space equator. In the case of the Bianchi spacetimes, these great circles represent Kasner solutions. As a result of the analytical treatment, we see that the Mixmaster behaviour of Bianchi IX cosmologies must be resolved to quiescence by the introduction of a suitable matter field (in this case a scalar field) in order for the equations of motion to satisfy the Picard-Lindelöf theorem near the initial singularity. In other words, Bianchi IX quiescence must be achieved in order for the dynamical system to have a unique, smooth continuation through the big bang. In the numerical solution examples given for quiescent Bianchi IX in section 4.4, we see explicitly the final Kasner epochs and Taub transitions before the system passes through the initial singularity.

At present it remains unclear how one might develop a quantisation regime for shape dynamics, but certainly a proper understanding of the classical theory is required first. In particular, it is encouraging that the initial singularity is able to be resolved at the classical level within the Shape Dynamics framework. One might also hope that an investigation of geodesics on the contact manifold sheds light on possible alterations to the singularity theorems within the context of shape dynamics.

\begin{appendices}

\section{Bianchi Numerical Solution Hamiltonians}
\label{A_H}

\begin{figure}[h]
    \centering
    \includegraphics[scale = 0.7]{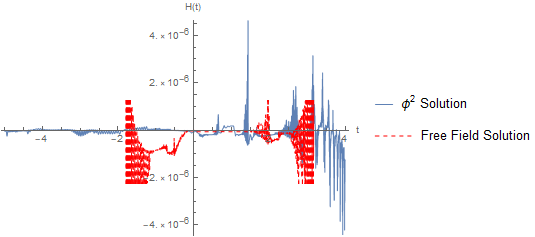}
    \caption{Numerical solution Hamiltonian for the Bianchi I + free field (red, dashed) and harmonic potential (blue) models. The discretised equations of motion for the free-field model become numerically faster than the harmonic potential model thus the numerical solution is valid over a smaller time domain.}
    \label{Bianchi_I_H}
\end{figure}

\begin{figure}[h]
\centering
\begin{subfigure}{0.45\textwidth}
    \includegraphics[width=\textwidth]{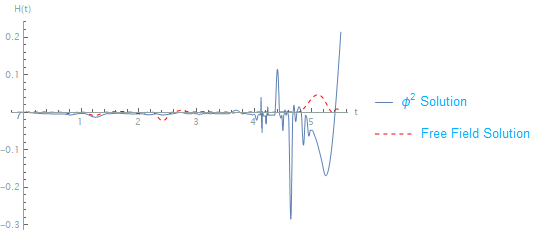}
    \caption{Bianchi IX numerical solution Hamiltonian for the free-field (red, dashed) and harmonic potential models. }
    \label{BIX_H}
\end{subfigure}
\hfill
\begin{subfigure}{0.45\textwidth}
    \includegraphics[width=\textwidth]{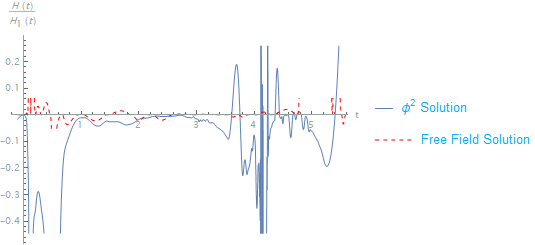}
    \caption{Plot of the ratio $\mathcal{H}/\mathcal{H}_1$ for the Bianchi IX numerical solution Hamiltonians.}
    \label{BIX_H_comp}
\end{subfigure}
\caption{}
    \label{}
\end{figure}

In figure \ref{BIX_H} we plot the numerical solution Hamiltonians for Bianchi IX, the numerical solution Hamiltonians are larger in magnitude than the previous cases we have looked at in this paper, however the Hamiltonian itself is still much smaller than the individual terms that it is comprised of. For example consider the first term in the Hamiltonian \ref{BIX_H_eq}
\begin{equation}
    \mathcal{H}_1 = \frac{1}{2}p_\beta^2\cos^2\beta
\end{equation}
The relative size of $\mathcal{H}/\mathcal{H}_1$ is typically small except where $\mathcal{H}_1 \rightarrow 0$, as demonstrated in figure \ref{BIX_H_comp}

\section{Gnomonic Projection Example}
\label{Gno}
Consider the motion of a free-particle in two dimensions, in the standard symplectic mechanics context. The Lagrangian for such a system is simply
\begin{equation}
    \mathcal{L} = \frac{1}{2}\left(\dot{x}^2 + \dot{y}^2\right)
\end{equation}
and the solutions, which are straight lines on the $\mathbb{R}^2$ plane are of course obtained trivially. The solutions are characterised by two constant momenta $(p_x,p_y)$ each associated with one of the coordinates on the plane
\begin{equation}
    x(t) = x_0 + p_xt, \quad y(t) = y_0 + p_yt
\end{equation}
In this appendix we show explicitly how, after making a gnomonic projection to shape space, the solutions of the equations of motion are great circles on $S^2$. Recall that the gnomonic projection is a compactification onto the surface of a sphere, in this case a 2-sphere. This projection takes a unit sphere tangent to the plane at the origin and draws a straight line from a point $P(x,y)$ on the plane to the centre of the tangent sphere. $P(x,y)$ is then identified with the coordinates on the 2-sphere $(\alpha,\beta)$ where the line intersects the sphere. The shape space coordinates are the angle subtended from the pole $\beta$ and the azimuthal angle $\alpha$.
Explicitly the projection map is written as 
\begin{equation}
    \begin{pmatrix}
        x \\
        y
    \end{pmatrix} = |\tan\beta|
    \begin{pmatrix}
        \cos\alpha \\
        \sin\alpha
    \end{pmatrix}
\end{equation}

In these shape space coordinates the Lagrangian in 
\begin{equation}
    \mathcal{L} = \frac{1}{2}\left(\dot{\beta}^2\sec^4\beta + \dot{\alpha}^2\tan^2\beta\right)
\end{equation}
From which one may find the canonical momenta 
\begin{equation}
    p_\alpha = \dot{\alpha}\tan^2\beta, \quad p_\beta = \dot{\beta}\sec^4\beta
\end{equation}
and obtain the Hamiltonian through a Legendre transformation
\begin{equation}
    \mathcal{H} = \frac{1}{2}p_\beta^2\cos^4\beta + \frac{p_\alpha^2}{2\tan^2\beta}
\end{equation}
One may now simply use the symplectic Hamiltons equations of motion \cite{deLeon:2020hnm}

\begin{equation}
    \dot{q}_i = \frac{\partial \mathcal{H}}{\partial p^i}, \quad \dot{p}^i = -\frac{\partial \mathcal{H}}{\partial q_i}
\end{equation}

to obtain
\begin{equation}
    \dot{\alpha} = \frac{p_\alpha}{\tan^2\beta}, \quad \dot{p}_\alpha = 0
\end{equation}
\begin{equation}
    \dot{\beta} = p_\beta\cos^4\beta, \quad \dot{p}_\beta = 2p_\beta^2\cos^3\beta\sin\beta + p_\alpha^2\frac{\cos\beta}{\sin^3\beta}
\end{equation}
In symplectic mechanics, the Hamiltonian is a conserved, constant quantity and therefore one may determine that
\begin{equation}
\label{Gno_eom}
    \frac{d\alpha}{d\beta} = \frac{\dot{\alpha}}{\dot{\beta}} = \frac{p_\alpha}{\sin^2\beta\sqrt{2\mathcal{H}-p_\alpha^2\cot^2\beta}}
\end{equation}

Equation \ref{Gno} has the solution
\begin{equation}
    \cos(\alpha-\alpha_0) = \frac{p_\alpha}{\sqrt{2\mathcal{H}}\tan\beta}
\end{equation}
which is the equation of a great circle on the 2-sphere, whose orientation relative to the $\mathbb{R}^2$ is controlled by $p_\alpha/\sqrt{2\mathcal{H}}$.

\end{appendices}

\bibliography{refs.bib}

\bibliographystyle{ieeetr}

\end{document}